\documentclass[11pt, a4paper]{article}
\usepackage{jheparxiv}
\usepackage[utf8]{inputenc}
\usepackage{amsmath}
\usepackage{bbold}
\usepackage{amsfonts}
\usepackage{amssymb}
\usepackage{latexsym}
\usepackage{mathrsfs}
\usepackage{braket}		
\usepackage{wasysym}
\usepackage{oplotsymbl}
\usepackage{graphicx}

\newcommand{\hCS}{\mathrm{hCS}}

\newcommand{\cc}{\mathsf{c}}

\newcommand{\gh}{\mathrm{gh}}

\newcommand{\sW}{\mathsf{A}}

\newcommand{\sB}{\mathsf{B}}

\newcommand{\Ind}{\mathrm{Ind}}

\newcommand{\Ccurl}{\mathscr{C}}

\newcommand{\sn}{\mathsf{n}}
\newcommand{\dd}{\mathsf{d}}

\newcommand{\Spec}{\mathtt{Spec}}

\newcommand{\nn}{\nonumber}
\newcommand{\Tod}{\mathtt{Tod}}
\newcommand{\Ch}{\mathtt{Ch}}

\newcommand{\V}{\mathbb{V}}
\newcommand{\sA}{\mathsf{A}}
\newcommand{\sF}{\mathsf{F}}

\newcommand{\rank}{\mathrm{rank}}

\usepackage{tikz-cd}
\usepackage{graphicx}
\usepackage{color}
\usepackage{xcolor}
\usepackage{slashed}
\usepackage{twistor}
\usepackage[all]{xy}
\usepackage{tikz-cd}
\usepackage{mathtools}
\usepackage{float}
\usetikzlibrary{arrows, matrix}
\tikzcdset{arrow style=math font}
\tikzset{>=latex}

\newcommand{\tc}{\mathtt{c}}

\newcommand{\varthetabold}{\boldsymbol{\vartheta}}

\newcommand{\tp}{\mathtt{p}}

\newcommand{\tH}{\mathtt{H}}

\usepackage{skak}

\newcommand{\Tr}{\text{Tr}}

\newcommand{\mg}{\mathfrak{g}}

\newcommand{\tW}{\boldsymbol{\mathsf{W}}}

\newcommand{\B}{\mathbb{B}}

\newcommand{\msp}{\mathfrak{sp}}
\newcommand{\eps}{\epsilon}

\newcommand{\msf}[1]{\mathsf{#1}}
\newcommand{\F}{\mathbb{F}}

\title{\Large Anomaly-free twistorial higher-spin theories}

\author[]{Tung Tran}
\affiliation[]{Asia Pacific Center for Theoretical Physics, POSTECH, \\
77 Cheongamro, Nam-gu,
Pohang-si, Gyeongsangbuk-do, 37673, Korea}
\emailAdd{tran.tung@apctp.org}

\abstract{We present twistor BV actions that encompass many classically consistent bosonic holomorphic twistorial higher-spin theories with vanishing cosmological constant. Upon quantization, these actions are shown to be quantum consistent, i.e. no gauge anomaly, for some subclasses of twistorial higher-spin theories. Anomaly-free twistorial theories can be identified through an index theorem, which is a higher-spin extension of the Hirzebruch-Riemann-Roch index theorem. 
We also discuss the anomaly cancellation mechanisms on twistor space to render anomalous theories quantum consistent at one loop. }

\begin{document}

\maketitle
\section{Introduction}
Twistor theory, cf. \cite{Penrose:1967wn}, is a fascinating subject, repeatedly found in some of the unexpected corners in mathematics and physics, connecting different topics/ideas together and some time can even unify them into a bigger framework. This has helped to deepen our understanding of existing structures as well as paving the way for the emergence of new theories and applications, with notable examples being the twistor string \cite{Witten:2003nn}, and the ambi-twistor string  \cite{Mason:2013sva,Geyer:2014fka}.

It is well-known that certain spacetime theories, such as self-dual Yang-Mills (SDYM) \cite{Chalmers:1996rq} or self-dual gravity (SDGR) \cite{Krasnov:2016emc,Krasnov:2021cva}, can be derived from holomorphic BF type theories on twistor space, cf. \cite{Mason:2005zm,Mason:2007ct}, which are integrable at classical level \cite{Ward:1977ta,Penrose:1976js}. This classical integrability can be traced back to the underlying $\infty$-dimensional chiral algebras \cite{Boyer:1985aj,Park:1989fz,Park:1989vq,Dunajski:2000iq}, which trivialize the tree-level scattering amplitudes of these theories. Recently, this idea has also been extended to non-commutative cases \cite{Monteiro:2022lwm,Monteiro:2022xwq,Bu:2022iak,Ponomarev:2022ryp}, where the introduction of higher-spin fields has been observed as a natural way to preserve Lorentz invariance.\footnote{See also e.g. \cite{Steinacker:2019awe,Steinacker:2023cuf,Steinacker:2024unq} for similar discussions in the context of a higher-spin theory induced by the IKKT matrix model.}


SDYM and SDGR are one-loop exact, i.e.  the only non-trivial amplitudes of these theories are the $n$-point one-loop amplitudes, cf. \cite{Bern:1993qk,Mahlon:1993fe,Cangemi:1996rx,Cangemi:1996pf,Bern:1998sv,Bern:1998xc}. These one-loop amplitudes are rational functions in the kinematics which contain only poles and have no branch cuts, a direct consequence of the triviality of tree-level amplitudes. Non-trivial one-loop amplitudes can then be interpreted as an ``anomaly'' that obstructs the conservation of the tower of infinitely many conserved charges associated with the integrability of self-dual theories \cite{Bardeen:1995gk}.\footnote{In fact, any theories with non-trivial amplitudes are referred to as \emph{not integrable} \`a la Bardeen \cite{Bardeen:1995gk}. Note that the word ``anomaly'' does not imply the $4d$ theories are pathological. Rather, it indicates the failure of a theory being quantum integrable at one loop by having non-trivial one-loop amplitudes.} 

To render SDYM and SDGR quantum integrable, i.e. to trivialize their one-loop amplitudes, one can introduce suitable axionic couplings \cite{Costello:2021bah, Bittleston:2022nfr} to the actions of SDYM and SDGR. These new axionic couplings generate tree-level diagrams with axions in the exchange, fine-tuning such that they can cancel the loop diagrams associated with the gauge anomaly. Note that the axionic couplings can be traced back to a Green-Schwarz-like mechanism of some holomorphic theories on twistor space, introduced to cancel the gauge anomaly of the corresponding twistorial theories. See also e.g. \cite{Costello:2015xsa, Costello:2019jsy} for previous work in this direction and \cite{Monteiro:2022nqt, Doran:2023cmj} for further discussion. 

Although the anomaly-cancellation mechanism in \cite{Costello:2021bah, Bittleston:2022nfr} is undeniably elegant, the introduction of additional degrees of freedom, i.e. the axion fields, will typically result in non-local 
spacetime actions when we integrate them out, challenging 
the fundamental principle of locality. As a result, it raises a compelling question: 
\begin{center}
    \emph{Do anomaly-free models exist without requiring such counterterms?}
\end{center}
We aim to address the above question by studying the quantization of various holomorphic BF and Chern-Simons type theories in twistor space by considering some parent twistor actions with gauge, gravitational and Moyal-Weyl-type couplings. 
For generality, we work in the language of the Batalin-Vilkovisky (BV) formalism \cite{Batalin:1981jr,Batalin:1983ggl} since it allows us to treat things uniformly. We show that there exist higher-spin theories that are anomaly-free at one loop. Moreover, these theories can be identified via an index derived from the Hirzebruch-Riemann-Roch index theorem adapted to the higher-spin case. 

The paper is organized as follows. Section \ref{sec:2} provides a brief review of twistor geometry and its inherent higher-spin symmetry. Section \ref{sec:3} presents the parent twistor BV actions for various holomorphic higher-spin twistor theories. Section \ref{sec:4} examines the quantization of the parent twistor BV actions at one loop. There, we point out various self-quantum consistent theories and discuss the anomaly cancellation mechanisms that render anomalous theories quantum consistent at one loop. We end with a discussion in Section \ref{sec:discuss}. There are also two appendices that are relevant to reading Section \ref{sec:4}.

\paragraph{Note added.} While finalizing this work, we became aware of complementary results by Lionel Mason and Atul Sharma \cite{Mason:2025pbz}, which may prove beneficial for future developments.

\section{Review}\label{sec:2}
This section reviews twistor geometry and the higher-spin symmetry it possesses. Unless otherwise stated, we will work with a complex setting, since the task of doing an analytic continuation to a Euclidean slice can be done with little subtlety. 
\subsection{Twistor geometry}\label{sec:twistor}
Let $\P^3=(\C^4-\{0\})/\C^*$ denote the $3d$ complex projective space with homogeneous coordinates $Z^A\sim t Z^A$ ($t\in \C^*$). Viewing $\P^3$ as the total space of the $\C^2$-bundle over $\P^1$ with the `element' $(Z^1,Z^2)=0$ removed, we can split (see e.g. for a review \cite{Adamo:2017qyl})
\begin{align}
   Z^A=(\lambda^{\alpha},\mu^{\dot\alpha})\,,\qquad A=1,2,3,4\,,\quad \alpha=1,2\,,\quad \dot\alpha=\dot{1},\dot{2}\,,
\end{align}
where $\lambda^{\alpha}$ are coordinates that describe points on the Riemann sphere $\P^1$ and $\mu^{\dot\alpha}$ are coordinates on the fiber. Note that $(\lambda^{\alpha},\mu^{\dot\alpha})$ can be viewed as left- and right-handed commutative spinors of the complexified Lorentz group $SO(4,\C)\simeq SL(2,\C)\times SL(2,\C)$. The \emph{undeformed} (projective) twistor space
\begin{align}
    \PT:=\big\{Z^A\in \P^3\,\big|\,\lambda^{\alpha}\neq 0\big\}\,,
\end{align}
is then defined as an open subset of $\P^3$ with the line $\lambda^{\alpha}=(Z^1,Z^2)=0$ removed. It will play the role of the background on which all types of deformations are considered. 

Let us denote $\cO_{\PT}(n)$ as the sheaf of local holomorphic functions with weight $n\in \N_0$ on $\PT$. Then, a section of $\cO_{\PT}(n)$ can be written locally as
\begin{align}
    f^{(n)}(Z)=f_{A(n)}Z^{A(n)}\,,\qquad Z^{A(n)}\equiv Z^{A_1}\ldots Z^{A_n}\,.
\end{align}
In terms of spinors $(\lambda^{\alpha},\mu^{\dot\alpha})$, a function $f^{(2s)}$ with weight $2s$ in $Z$ can be expressed as
\begin{align}
    f^{(2s)}(\lambda,\mu)=\sum_{m+n=2s}f_{\alpha(m)\,\dot\alpha(n)}\lambda^{\alpha(m)}\mu^{\dot\alpha(n)}\,,\quad \lambda^{\alpha(m)}=\lambda^{\alpha_1}\ldots\lambda^{\alpha_m}\,,\quad \text{etc.}
\end{align}
In what follows, we shall impose the incidence relations \cite{kodaira1962theorem,kodaira1963stability,Penrose:1967wn}:
\begin{align}\label{eq:incidence1}
    \mu^{\dot\alpha}=\mu^{\dot\alpha}(x,\lambda)\,, \qquad x^{\alpha\dot\alpha}:=x^{a}\sigma_{a}^{\alpha\dot\alpha}\,,\qquad a=1,2,3,4\,,
\end{align}
where $\sigma_a$ are quartenions. These relations tell us that the spinors $\mu^{\dot\alpha}$ are parametrized by a 4-dimensional moduli space $\cM$ with coordinates $x^{a}$, which may be identified with a 4-dimensional complexified spacetime. 

Observe that upon imposing the incidence relations, the sheaves of local holomorphic functions 
\begin{align}
    \cO_{\P^1}(n):=\{f(x,\lambda)\in\cO_{\P^1}(n)\,, \forall t\in\C^*\big|f(x,t\lambda)=t^nf(x,\lambda)\}\,,    
\end{align}
can serve as a generating space for many fundamental building blocks 
on $\PT$. Therefore, $\PT$ can be referred to as the total space of the bundle $\cO_{\P^1}(1)\oplus\cO_{\P^1}(1)\rightarrow\P^1$.


Being a subset of $\C^4$, it is clear that $SU(4)$ has a natural action on $\PT$. In particular, it induces a quaternionic conjugation, which can map $Z$, equivalently $(\lambda,\mu)$, to their $\mathbb H$-conjugate counterparts as
\begin{align}
    Z^A\mapsto \hat Z^A=(\hat\lambda^{\alpha},\hat\mu^{\dot\alpha})\,,\qquad \hat\lambda^{\alpha}=(-\overline{\lambda^2},\overline{\lambda^1})\,,\quad \hat\mu^{\dot\alpha}=(-\overline{\mu^{\dot 2}},\overline{\mu^{\dot 1}})\,.
\end{align}
This leads to the notation of sheaves of local functions on $\PT$ with weight $(m,n)$ in $(\lambda,\hat\lambda)$ denoted as $\cO_{\P^1}(m,n)$. As a vector space
\begin{align}\label{eq:weighted-spinors}
    \cO_{\P^1}(m,n):=\Big\{f(x,\lambda,\hat\lambda)\in\cO_{\P^1}(m,n)\,\Big|\,f(t\lambda,t^{-1}\hat\lambda)=t^{m-n}f(x,\lambda,\hat\lambda)\Big\}\,.
\end{align}
Here, the \emph{weight} or homogeneity of a function $f\in \cO_{\PT}(m,n)$ can be measured by the following operator
\begin{align}\label{eq:weight-operator}
    \tW:=\lambda^{\alpha}\p_{\alpha}
    -\hat\lambda^{\alpha}\hat\p_{\alpha}\,,
\end{align}
where
\begin{align}
    \p_{\alpha}\equiv\frac{\p}{\p\lambda^{\alpha}}\,,\qquad \hat\p_{\alpha}\equiv \frac{\p}{\p\hat\lambda^{\alpha}}\,.
\end{align}
By definition, $\lambda$ has weight one and $\hat\lambda$ has weight minus one. 

Notice that a function valued in $\cO_{\P^1}(m,n)$ has weight $m-n$ but spin $s=\frac{m+n}{2}$. Thus, to appropriately describe a massless spin-$s$ field with negative helicity, we may consider the following sheaves of holomorphic line bundle
\begin{align}
    \cO_{\P^1}(-n):=\Big\{f_{-n}(x,\lambda,\hat\sigma)\in \cO_{\P^1}(-n)\,\Big|\,f_{-n}(x,t\lambda,t^{-1}\hat\sigma)=t^{-n}f(x,\lambda,\hat\sigma)\Big\}\,,
\end{align}
where $\langle\lambda\,\hat\lambda\rangle\equiv \lambda^{\alpha}\hat\lambda_{\alpha}$ and $\hat\sigma^{\alpha}\equiv\frac{\hat\lambda^{\alpha}}{\langle\lambda\,\hat\lambda\rangle}$. (See below for the convention we use in this work.) This allows us having a unified treatment for weight and helicity of a massless particle, see e.g. \cite{Eastwood:1981jy,Mason:2007ct,Bullimore:2011ni,Bittleston:2020hfv} for previous work. 

Hereinafter, we will drop the subscript $\P^1$ in $\cO_{\P^1}(2h-2)$ for $h\in \Z$ for readability. Furthermore, as we want to work entirely on twistor space, it is more convenient to work with the $U(1)$-bundle over $\PT$, which has the geometry of an $S^7$. This allows us to identify the weights of $(\lambda,\mu,\hat\lambda,\hat\mu)$ with the $U(1)$-charges. In particular, as twistor variables are charged under $U(1)\subset \C^*$, their monodromy around $S^1$ is completely specified by the phase $e^{i\theta w}$ where $\theta\in [0,2\pi]$ and $w$ is the weight or charge associated to each twistor variable. As a result, we may define 
\begin{align}
    \cO(n):=\Big\{f_n(\lambda,\hat\sigma,\mu,\hat q)\in \cO(n)\big|f_n(t\,\lambda,t^{-1}\hat\sigma,t\,\mu,t^{-1}\,\hat q)=t^nf(\lambda,\hat\sigma,\mu,\hat q)\Big\}\,,
\end{align}
where
\begin{align}
    n\in \Z\,,\qquad \hat \sigma^{\alpha}=\frac{\hat\lambda^{\alpha}}{\langle\lambda\,\hat\lambda\rangle}\,,\quad \hat q^{\dot\alpha}=\frac{\hat\mu^{\dot\alpha} }{[\mu\,\hat\mu]}\,,
\end{align}
understanding $\mu=\mu(x,\lambda,\hat\lambda)$ and $\hat\mu=\hat\mu(x,\lambda,\hat\lambda)$. Consequently, this leads to the following notion of sheaves of holomorphic line bundle
\begin{align}\label{eq:master-line-bundle}
    \cO(2h-2):=\Big\{f_{2h-2}\in \cO(2h-2)\ \text{for}\ h\in \Z\,\Big|\,\tW f_{2h-2}=(2h-2)f_{2h-2}\Big\}\,.
\end{align}


\subsection{Higher-spin symmetry}  
This subsection reviews the holomorphic higher-spin symmetry of twistor space denoted as h$\hs_{\Lambda}$ where $\Lambda$ is the cosmological constant of the moduli space $\cM$. 
\paragraph{Holomorphic $\star$-product.} To construct various self-dual/chiral theories on twistor space, let us first introduce the holomorphic $\star$-product:
\begin{align}\label{eq:star-product}
    f_1(Z)\star f_2(Z)&= \exp\Big(\Pi_{1,2}\Big) f_1(Z_1)\,f_2(Z_2)\nn\\
    &=\exp\Big(\Lambda \langle\p_1\,\p_{2}\rangle+[\p_1\,\p_2]\Big)f_1(\lambda_1,\mu_1)f_2(\lambda_2,\mu_2)\Bigg|_{\substack{\lambda_{1,2}=\lambda\\
    \mu_{1,2}=\mu}}\,,
\end{align}
which is 
an associative Moyal-Weyl deformation of the Poisson bracket, cf. \cite{Fletcher:1990ib},
\begin{align}\label{eq:Poisson-structure}
    \Pi_{1,2}=-I^{AB}\frac{\p}{\p Z_1^A}\frac{\p}{\p Z_2^B}\,,\qquad  I^{AB}=\begin{pmatrix} \Lambda \epsilon^{\alpha\beta} & 0 \\
    0 & \epsilon^{\dot\alpha\dot\beta}\end{pmatrix}\,,\quad I_{AB}=\begin{pmatrix} \epsilon_{\alpha\beta} & 0 \\
    0 & \Lambda \epsilon_{\dot\alpha\dot\beta}\end{pmatrix}\,.
\end{align}
Here, $I^{AB}$ is the infinity twistor \cite{Penrose:1967wn} obeying $I_{AC}I^{BC}=\Lambda \delta_A{}^B$. This is an object that defines the geometry at infinity, as well as introduces a metric structure into various class of gravitational twistorial theories. Note that
\begin{align}
     \langle \p_i\,\p_j\rangle=\p_i^{\alpha}\p_{j\alpha}\,,\qquad  [\p_i\,\p_j]=\p_{i}^{\dot \alpha}\p_{j\dot \alpha}\,;\qquad \p_{i\alpha}\equiv\frac{\p}{\p\lambda_i^{\alpha}}\,,\quad \p_{i\dot\alpha}\equiv\frac{\p}{\p\mu_i^{\dot\alpha}}\,.
\end{align}
Our convention for raising and lowering  spinorial indices is
\begin{align}
    v^{\alpha}=\epsilon^{\alpha\beta}v_{\beta}\,,\quad v_{\alpha}=v^{\beta}\epsilon_{\beta\alpha}\,,\quad v^{\dot\alpha}=\epsilon^{\dot\alpha\dot\beta}v_{\dot\beta}\,,\quad v_{\dot\alpha}=v^{\dot\beta}\epsilon_{\dot\beta\dot\alpha}\,,
\end{align}
where $\eps^{\alpha\beta}$ is the $\msp(2)$-invariant tensor with the properties $\eps^{\alpha\beta}=\eps_{\alpha\beta}=-\eps^{\beta\alpha}$ and $\eps^{12}=1$. Observe that the Weyl algebra $\sW_2$ of polynomial functions in $Z^A=(\lambda^{\alpha},\mu^{\dot\alpha})$ defined in terms of the holomorphic $\star$-product cf. \eqref{eq:star-product} can be identified as the higher-spin algebra 
\begin{align}
    \mathrm{h}\hs_{\Lambda}:=\sA_1(\lambda)\otimes  \sW_1(\mu)\,.
\end{align}
Here, the subscripts in $\sA$ denote the number of canonical pairs.

\section{Parent twistor BV actions}\label{sec:3}
This section proposes parent twistor actions which can give rise to various chiral/self-dual theories in spacetime. For various technical issues related to the holomorphic measure on twistor space, see e.g. discussion in \cite{Adamo:2013tja,Haehnel:2016mlb,Adamo:2016ple,Tran:2022tft}, we are only able to provide actions that work for the case of vanishing cosmological constant. In this case, the Poisson structure \eqref{eq:Poisson-structure}, and the holomorphic $\star$-product 
reduce to 
\begin{align}
    \Pi_{1,2}^{\Lambda=0}=[\p_1\,\p_2]\,,\quad \star =\exp\Big(\Pi_{1,2}^{\Lambda=0}\Big)\,,
\end{align}
\normalsize
Henceforth, we suppress the superscript notation $\Lambda=0$ as no confusion can arise.

\subsection{Actions for chiral/self-dual higher-spin theories} 
As usual, to construct an anti-holomorphic Lagrangian $(0,3)$-form of weight $-4$, which is integrated against the canonical holomorphic measure 
\begin{align}
    \Omega^{(3,0)}:=\langle \lambda \,d\lambda\rangle\wedge [d\mu\wedge d\mu]
    \,,\qquad \tW\,\Omega^{(3,0)}=4\,,
\end{align}
on $\PT$, one must keep track of the spinor weights of the field content to make sure that the action is weightless in $\lambda$ and $\hat{\lambda}$. That is, any twistor action on $\PT$ must be an element of $\Omega^{3,3}(\PT,\cO(0))$. We remind the reader that $\mu^{\dot\alpha}=\mu^{\dot\alpha}(x,\lambda)$ is of weight one in $\lambda$ by virtue of the incidence relations \eqref{eq:incidence1} and the weight operator $\tW$, cf. \eqref{eq:weight-operator}.


As alluded to in the above, it will be more convenient to construct twistor actions on $S^7=\PT\times U(1)$ and consider a projection to the base $\PT$. Since we will do an angular integral to project our actions on $S^7$ down to $\PT$, only expressions with trivial monodromy will survive.

By definition, the Poisson structure has charge $-2$ under the action of $\tW$, cf. \eqref{eq:weight-operator}, and can be written as 
\begin{align}\label{eq:Poisson-theta}
    \Pi(-,-)\mapsto\Pi_{\theta}(-,-)=e^{-2i\theta}[\p_i\,\p_j]\,,\qquad \p_{i\dot\alpha}=\frac{\p}{\p\mu_i^{\dot\alpha}}\,.
\end{align}
\subsubsection{Parent twistor actions for chiral theories}
The parent twistor actions for $4d$ chiral theories \cite{Metsaev:1991mt,Metsaev:1991nb,Ponomarev:2016lrm,Ponomarev:2017nrr} have the following simple forms:\footnote{See also  \cite{Mason:2025pbz} for complementary results.}
\begin{subequations}\label{eq:parent-action-1}
    \begin{align}
    S_{hCS_1}^{[,]_{\mg}}&=\int_{S^7}d\theta \,\Omega^{3,0} \,\Tr\Big(\sA\wedge\bar{\p}\sA+\frac{1}{3}\sA\wedge [\sA,\sA] \Big)\,,\label{eq:BV-Lie}\\
    S_{hCS_2}^{\{,\}}&=\int_{S^7}d\theta \,\Omega^{3,0} \,\Big(\sA\wedge\bar{\p}\sA+\frac{1}{3}\sA\wedge\{\sA, \sA\} \Big)\,,\label{eq:BV-Poisson}\\
    S_{hCS_3}^{\star}&=\int_{S^7}d\theta \,\Omega^{3,0} \,\Tr\Big(\sA\star\bar{\p}\sA+\frac{2}{3}\sA\star \sA \star \sA \Big)\,,\label{eq:BV-star}
\end{align}
\end{subequations}
where 
$\bar{\p}$ is the usual Dolbeault operator on $\P^3$ defined by 
\begin{align}
    \bar{\p}:=d\hat Z^A\frac{\p}{\p \hat{Z}^A}= d\hat\lambda^{\alpha}\hat{\p}_{\alpha}+ d\hat{\mu}^{\dot\alpha}\,\hat{\p}_{\dot\alpha}\,,\qquad \hat\p_{\alpha}=\frac{\p}{\p\hat\lambda^{\alpha}}\,,\quad \hat\p_{\dot\alpha}=\frac{\p}{\p\hat\mu^{\dot\alpha}}\,.
\end{align}
Here, we have indexed the twistor actions with the kind of interactions they possess. Namely, the usual Lie algebra commutation relation $[\,,]_{\mg}$, the Poisson bracket $\{\,,\}$, and the $\star$-product. It is also useful to note that the $\star$-product, which generates higher-derivative interactions in $S_{hCS_3}^{\star}$ is chosen such that it does not deform the holomorphic measure $\Omega^{(3,0)}$ on twistor space.\footnote{Had we consider the Poisson structure associated with the infinity twistor \eqref{eq:Poisson-structure} with $\Lambda\neq 0$, the measure will deform leading to a non-invariant action as noticed earlier in \cite{Tran:2022tft}. To construct a gauge invariant action in this case, one may need to resort to the construction of trace invariants \cite{feigin2005hochschild} adapted to Fedosov's framework \cite{Fedosov:1994zz} with a holomorphic twist \cite{kapranov1999rozansky}.} 

Let us now discuss the field content in the actions \eqref{eq:parent-action-1}. In the case of $S^{[,]_{\mg}}_{hCS_1}$, the tower of higher-spin fields can be encoded by a $\mg$-valued weighted higher-spin generating connection $(0,1)$-form
\begin{align}\label{eq:A-generating-function}
    \sA=\sum_{h\in \Z}e^{i\theta(2h-2)}\sA_{2h-2}\,,\qquad \sA_{2h-2}\in \Omega^{0,1}(\PT,\cO(2h-2)\otimes \mg)\,,
\end{align}
where $\mg$ is some Lie algebra, and $\sA_{2h-2}$ are $(0,1)$-forms twisted by $\cO(2h-2)$. Here, $\sA_{2h-2}$ are some local functions in $(x,\lambda,\hat\lambda)$ (or equivalently $(\lambda,\hat\lambda,\mu,\hat\mu)$ with the incidence relations imposed) with the total $U(1)$-charge\footnote{Although these discrete states look like Kaluza-Klein modes, all higher-spin fields are, nevertheless, massless in our setting. This can be seen at the linearized level when the incidence relations are simple enough to solve. See e.g. \cite{Adamo:2022lah}.}  equal to $2h-2$ for $h\in \Z$. 

On the other hand, if we consider $S_{hCS_2}^{\{,\}}$, it is necessary that $\mg=\mathfrak{u}(1)$. In other words, the higher-spin generating connection is colorless in this case. For the case of $S_{hCS_3}^{\star}$, the higher-spin connection can be either colored or colorless.

It can be checked that the actions \eqref{eq:parent-action-1} are gauge invariant under
\begin{subequations}
    \begin{align}
      S^{[,]_{\mg}}_{hCS_1}&: \quad  &\delta_{\xi} \sA&=\bar{\p}\xi+[\sA,\xi]\,,\label{eq:gauge-variation-Lie}\\
      S^{\{,\}}_{hCS_2}&: \quad  &\delta_{\xi} \sA&=\bar{\p}\xi+\{\sA,\xi\}\,,\label{eq:gauge-variation-Poisson}\\
      S^{\star}_{hCS_3}&: \quad  &\delta_{\xi} \sA&=\bar{\p}\xi+[\sA,\xi]_{\star}\,.
    \end{align}
\end{subequations}
where
\begin{align}
    \xi=\sum_{h\in \Z}e^{i\theta(2h-2)}\xi_{2h-2}\,,\qquad \xi_{2h-2}\in\Gamma(\PT,\cO(2h-2)\otimes\mg)\,,
\end{align}
for $S_{hCS_1}^{[,]}$ and $S_{hCS_3}^{\star}$; and $\mg=\mathfrak{u}(1)$ for $S_{hCS_2}^{\{,\}}$ case. Note that the last action in \eqref{eq:parent-action-1} can be simplified by doing integration by part. In particular, we get
\begin{align}\label{eq:parent-action1.5}
    S_{hCS_3}^{\star}=\int_{S^7}d\theta \,\Omega^{3,0} \,\Tr\Big(\sA\wedge\bar{\p}\sA+\frac{2}{3}\sA \wedge\sA \star \sA \Big)\,,
\end{align}
thanks to the fact that $\p^{\dot\alpha}\p_{\dot\alpha}\sA=0$, cf. \cite{Tran:2022tft}. As mentioned in footnote 4, the integration by part do not generate additional terms when $\p_{\dot\alpha}$ acts on the measure  $\Omega^{3,0}$, which would otherwise compromise gauge invariance, as in the case of $\Lambda\neq 0$ (see \cite{Adamo:2013tja,Haehnel:2016mlb,Adamo:2016ple,Tran:2022tft}). 


\paragraph{Twistor actions on the base.} To obtain the suitable twistor actions on the base, we substitute \eqref{eq:A-generating-function} and \eqref{eq:Poisson-theta} to \eqref{eq:parent-action-1}, and doing integrations over the angular variable $\theta$. Note that the actions on the total space $S^7$ must have trivial monodromy, or neutral $U(1)$ charge, in order to induce non-trivial twistor actions on the base. We refer to this sequence of operations as the pushforward to the base. 
Henceforth, we shall suppress the wedge products for readability.

It is easy to notice that, at free level, all twistor actions in \eqref{eq:parent-action-1} have the same kinetic term:
\begin{align}\label{eq:kinetic-action}
    S_{kin}=\int_{\PT}\Omega^{3,0}\Tr\Big(\sum_h\sA_{-2|h|-2}\bar{\p}\sA_{2|h|-2}\Big)\,,\qquad h\in \Z\,.
\end{align}
Now, at the level of interactions, we shall require the fields entering the vertices obey only the trivial monodromy constraint on the total space $S^7$. This leads to various holomorphic Chern-Simons theories.

$\bullet$ \underline{$S^{[,]_{\mg}}_{hCS_1}$ on the base.} Let us first proceed with the action $S^{[,]_{\mg}}_{hCS_1}$. Its holomorphic Chern-Simons action on the base reads
\begin{align}\label{eq:hCS-Lie-1}
    S_{hCS_1}^{[\,,]_{\mg}}=\int_{\PT}\Omega^{3,0}\Tr\Big(\sum_h\sA_{-2|h|-2}\bar{\p}\sA_{2|h|-2}+\frac{1}{3}\sum_{\{h_i\in \Spec\}}\sA_{2h_1-2}[\sA_{2h_2-2},\sA_{2h_3-2}]\Big)\,.
\end{align}
Observe that even though this action is somewhat similar to the action of holomorphic Chern-Simons in \cite{Adamo:2022lah}. However, we do not need to implement specific choice of boundary conditions on the $\Omega^{3,0}$ form to make \eqref{eq:hCS-Lie-1} well-defined. Furthermore, the reduction from twistor space to spacetime in this case is simple thanks to the fact that the incidence relations \eqref{eq:incidence1} can be solved explicitly by imposing the holomorphicity condition $\bar{\p}\mu^{\dot\alpha}=0$. In particular, we get $\mu^{\dot\alpha}=x^{\alpha\dot\alpha}\lambda_{\alpha}$ in this case, assuming the deformation $[\sA,-]$ is sufficiently small so that the topology of $\PT$ remains trivial. 

From the monodromy constraint, we can read off the helicity condition
\begin{align}
    h_1+h_2+h_3=1\,.
\end{align}
This means that beside the choice of having the spectrum $\Spec=\Z$, one can also have the option where $\Spec=2\Z+1$, cf. \cite{Monteiro:2022xwq}. Here, $\Spec$ refers to the space of fields with different helicities. \\
$\bullet$ \underline{$S^{\{,\}}_{hCS_2}$ on the base.} Let us proceed with the action \eqref{eq:BV-Poisson}. Analogously with the above we obtain
\begin{align}
    S_{hCS_2}^{\{\,,\}}&=\int_{\PTc}\Omega^{3,0}\Tr\Big(\sum_h\sA_{-2|h|-2}\bar{\p}\sA_{2|h|-2}+\frac{1}{3}\sum_{\{h_i\}}\sA_{2h_1-2}\{\sA_{2h_2-2},\sA_{2h_3-2}\}\Big)\,.
\end{align}
The helicity constraint and the spin constraint read
\begin{align}
    h_1+h_2+h_3&=2\,.
\end{align}
Here, besides $\Spec=\Z$, we can also have $\Spec=|h|\geq 2$ or $\Spec=2\Z$. Each case leads to a different conclusion when we quantize the corresponding theories at one loop.

Note that the Hamiltonian vector flow $\{\sA_2,-\}$, which can be viewed as a vector-valued $(0,1)$-form aka. a Beltrami differential, will deform $\PT$ to $\PTc$ -- the curved or deformed twistor space \cite{Penrose:1976js}. In this case, the incidence relations are harder to solve, as the covariantly constant condition:
\begin{align}
    \bar{\p}\mu^{\dot\alpha}+\{\sA_2,\mu^{\dot\alpha}\}=0\,,
\end{align}
is a non-linear pde, which may look deceptively simple. For this reason, in the remaining of this work, we will stay on twistor space and leave the reduction to spacetime for a future work.\\
$\bullet$ \underline{$S^{\star}_{hCS_3}$ on the base.} Writing the $\star$-product as $\sum_k\frac{1}{k!}\Pi^k$ and restricting ourselves to the trivial monodromy sector on $S^7$, we obtain the following actions on $\PTc$
\begin{align}\label{eq:parent-action-2-hCS}
    S^{\star}_{hCS_3}&=\int_{\PTc}\Omega^{3,0}\Tr\Big(\sum_h\sA_{-2|h|-2}\bar{\p}\sA_{2|h|-2}+\frac{2}{3}\sum_{\{h_i\}}\frac{1}{k!}\sA_{2h_1-2}\Pi^k(\sA_{2h_2-2},\sA_{2h_3-2})\Big)\,.
\end{align}
The helicity constraints can be expressed via the well-posedness of the couplings. We have
\begin{align}
    k=h_1+h_2+h_3-1\,,\label{eq:helicity-constraint-star}
\end{align}

\paragraph{Summary.} Below, we summarize the descendant theories of \eqref{eq:parent-action-1}, for the cases where $\sA$ is $\mg$-valued, and where $\sA$ is colorless.

$\bullet$ \underline{Colored holomorphic CS theories.} Let us first consider the colored case, where $\sA$ is given by $\sA\equiv \sA_aT^a$ with $T^a$ the generators of certain non-abelian Lie algebra $\mg$. In this case, the Lie bracket and holomorphic $\star$-product can be written explicitly as
\begin{subequations}
    \begin{align}
    [\sA,\sA]&=T_cf^{abc}\sA_a\sA_b\,,\\
    \sA\star \sA&= \sum_kT_c\,\frac{\tg^{abc}_{k}}{k!}\Pi^k(\sA_a,\sA_b)\,,\qquad k=h_1+h_2+h_3-1\,.
\end{align}
\end{subequations}
Here, $\tg^{abc}_k$ is a structure constant which depends on the number of derivatives. In particular,
\begin{align}
    \tg_{k\in 2\N_0}^{abc}=f^{abc}\,,\qquad \tg_{k\in 2\N-1}^{abc}=d^{abc}\,,
\end{align}
where $f^{abc}$ and $d^{abc}$ are the structure constants obtained from the usual Lie algebra relations  $[T^a,T^b]=f^{abc}T_c$ and $\{T^a,T^b\}_+=d^{abc}T_c$. The descendant Lorentz invariant theories from the parent action \eqref{eq:parent-action-1} are
\begin{table}[h!]
    \centering
    \begin{tabular}{|c|c|c|}\hline
       Theory  & spectrum & interaction type   \\ \hline\hline
       $\hCS^{\star}_{\mg}$ & all spins & $\star $ \\\hline
       $\hCS_{\mg}$ & all/or odd spins & $[\,,]_{\mg}$\\\hline
    \end{tabular}
    \caption{Table of descendant \underline{\emph{colored}} higher-spin theories derived from the parent actions \eqref{eq:parent-action-1}.}
    \label{tab:a}
\end{table}

$\bullet$ \underline{Colorless holomorphic CS theories.} When $\sA$ is $\mathfrak{u}(1)$-valued, there is no ambiguity in handling the $\star$-product. We obtain the following theories
\begin{table}[h!]
    \centering
    \begin{tabular}{|c|c|c|c|}\hline
       Theory  & spectrum & interaction type  \\ \hline\hline
       $\hCS^{\star}$ & all & $\star $  \\\hline
       $\hCS^{\{,\}}$ & all/or even+0 & $\{\,,\}$\\\hline
    \end{tabular}
    \caption{Table of descendant \underline{\emph{colorless}} higher-spin theories derived from the parent action \eqref{eq:parent-action-1}.}
    \label{tab:b}
\end{table}

\subsubsection{Parent twistor actions for self-dual theories}
For spacetime self-dual theories, we have the following parent twistor actions
\begin{subequations}\label{eq:BF-parents}
    \begin{align}
        S_{BF_1}^{[,]_{\mg}}&=\int_{S^7}d\theta \,\Omega^{3,0} \,\Tr\Big(\cB\bar{\p}\cA+\frac{1}{2}\cB[\cA,\cA] \Big)\,,\label{eq:BF-Lie}\\
    S_{BF_2}^{\{,\}}&=\int_{S^7}d\theta \,\Omega^{3,0} \,\Big(\cB\bar{\p}\cA+\frac{1}{2}\cB\{\cA, \cA\} \Big)\,,\label{eq:BF-Poisson}
    \end{align}
\end{subequations}
where we note that there is no option of having a $\star$-product in these cases while maintaining Lorentz invariance. In particular, any attempt to introduce the $\star$ product will force the above theories to become $\hCS^{\star}$.

In the case of $S_{BF_1}^{[,]_{\mg}}$, we have
\begin{subequations}\label{eq:generating-BF}
    \begin{align}
        \cA&=\sum_{s\in \N}e^{i\theta(2s-2)}\cA_{2s-2}\,,\qquad &\sA_{2s-2}&\in\Omega^{0,1}(\PT,\cO(2s-2)\otimes\mg)\,,\\
        \cB&=\sum_{s\in \N}e^{i\theta(-2s-2)}\cB_{-2s-2}\,,\qquad &\sB_{-2s-2}&\in\Omega^{0,1}(\PT,\cO(-2s-2)\otimes\mg)\,,
    \end{align}
\end{subequations}
while in the case of $S_{BF_2}^{\{,\}}$, it is necessary that $\mg=\mathfrak{u}(1)$, i.e. fields are colorless. 

Although, one may think of these BF-type theories as some closed subsectors of the holomorphic Chern-Simons theories \eqref{eq:parent-action-1} with some of the vertices deleted, it is not clear how these theories can be related since the gauge transformations of the positive-helicity fields $\cA$ and negative-helicity fields $\cB$ are slightly different from those in \eqref{eq:gauge-variation-Lie} and \eqref{eq:gauge-variation-Poisson}. In particular,
\begin{subequations}\label{eq:gauge-variation-BF}
    \begin{align}
        \text{In}\  S^{[,]_{\mg}}_{BF_1}&: \quad  &\delta_{\xi} \cA&=\bar{\p}\xi+[\cA,\xi]\,,\qquad  &\delta_{\xi,\chi}\cB&=\bar{\p}\chi+[\cA,\chi]+[\cB,\xi]\,,\label{eq:gauge-variation-Lie-BF}\\
     \text{In}\  S^{\{,\}}_{BF_2}&: \quad  &\delta_{\xi} \cA&=\bar{\p}\xi+\{\cA,\xi\}\,,\qquad &\delta_{\xi,\chi}\cB&=\p\chi+\{\cA,\xi\}+\{\cB,\xi\}\,,\label{eq:gauge-variation-Poisson-BF}
    \end{align}
\end{subequations}
where for $S^{[,]_{\mg}}_{BF_1}$
\begin{subequations}
    \begin{align}
    \xi&=\sum_{s\in \N}e^{i\theta(2s-2)}\xi_{2s-2}\,,\qquad &\xi_{2s-2}&\in\Gamma(\PT,\cO(2s-2)\otimes\mg)\,,\\
    \chi&=\sum_{s\in \N}e^{i\theta(-2s-2)}\xi_{-2s-2}\,,\qquad &\chi_{-2s-2}&\in\Gamma(\PT,\cO(-2s-2)\otimes\mg)\,.
\end{align}
\end{subequations}
Here, we choose the Lie algebra $\mg$ is such that it is isomorphic to its dual $\mg^{\vee}$. In the case of $S^{\{,\}}_{BF_2}$, it is necessary that $\mg=\mathfrak{u}(1)$.

\paragraph{Twistor actions on the base.} To obtain the suitable twistor actions on the base, we substitute \eqref{eq:generating-BF} and \eqref{eq:Poisson-theta} to \eqref{eq:BF-parents}, and doing integrations over the angular variable $\theta$. We obtain: 

$\bullet$ \underline{$S^{[,]_{\mg}}_{BF_1}$ on the base.} In the case of $S_{hBF_1}^{[,]_{\mg}}$, the twistor action on the base is
\begin{align}\label{eq:hBF-Lie-1}
    S_{hBF_1}^{[,]_{\mg}}=\int_{\PT}\Omega^{3,0}\Tr\Big(\sum_s\cB_{-2s-2}\bar{\p}\cA_{2s-2}+\frac{1}{2}\sum_{\{s_i\}}\cB_{-2s_1-2}[\cA_{2s_2-2},\cA_{2s_3-2}]\Big)\,.
\end{align}
In this case, we obtain the following spin constraint
\begin{align}
    s_2+s_3=s_1+1\,,
\end{align}
where it is possible to have
\begin{align}
    \Spec=\Z^*\,,\qquad \text{or}\qquad \Spec=2\Z+1\,.
\end{align}
It is worth noting that the spacetime dual theory of \eqref{eq:hBF-Lie-1} is known as self-dual higher-spin Yang Mills theory \cite{Krasnov:2021nsq,Herfray:2022prf,Adamo:2022lah}. 

$\bullet$ \underline{$S^{\{,\}}_{BF_2}$ on the base.} In the case of $S^{\{,\}}_{BF_2}$, the twistor action on the base is
\begin{align}
    S_{hBF_2}^{\{\,,\}}=\int_{\PTc}\Omega^{3,0}\Big(\sum_s\cB_{-2s-2}\bar{\p}\cA_{2s-2}+\frac{1}{2}\sum_{\{s_i\}}\cB_{-2s_1-2}\{\cA_{2s_2-2},\cA_{2s_3-2}\}\Big)
\end{align}
where the spin constraint reads
\begin{align}
    -s_1+s_2+s_3=2\,.
\end{align}
Note that even though it may be tempting to write down also a BF-type theory with Moyal-Weyl $\star$-product with the spin constraint $ m=-s_1+s_2+s_3-1$. It turns out that we will need the scalar field with $s=0$ in the spectrum to make this theory consistent. Thus, these type of theories will be forced to become $S^{\star}_{hCS_3}$, as alluded to in the above. 

\paragraph{Summary.} Below, we summarize the descendant theories of \eqref{eq:BF-parents} based on their color.

\begin{table}[ht!]
    \centering
    \begin{tabular}{|c|c|c|}\hline
       Theory  & spectrum & interaction type   \\ \hline\hline
       HS-BF${}_{\mg}$ & $s\geq 1$& $ [\,,]_{\mg}$  \\\hline
       HS-BF${}_{\mg}$ & odd spins & $ [\,,]_{\mg}$  \\\hline
    BF${}_{\mg}$ & $s=1$ & $[\,,]_{\mg}$  \\\hline
    \end{tabular}
    \caption{Table of descendant \underline{\emph{colored}} higher-spin theories derived from the parent actions \eqref{eq:BF-parents}.}
    \label{tab:BF-a}
\end{table}

\begin{table}[ht!]
    \centering
    \begin{tabular}{|c|c|c|c|}\hline
       Theory  & spectrum & interaction type  \\ \hline\hline
       HS-BF${}_{GR}$ & $s\geq 2$ & $ \{\,,\}$  \\\hline
       HS-BF${}_{GR}$ & $s\in 2\N$ & $ \{\,,\}$  \\\hline
       BF${}_{GR}$ & $s=2$ & $\{\,,\}$  \\\hline
    \end{tabular}
    \caption{Table of descendant \underline{\emph{colorless}} higher-spin theories derived from the parent action \eqref{eq:BF-parents}.}
    \label{tab:BF-b}
\end{table}



\subsection{BV-BRST formalism for holomorphic twistorial higher-spin theories}
In the previous subsection, we have shown that in the case where the infinity twistor is degenerate (with $\Lambda=0$), many twistor actions can be written in terms of the Chern-Simons $(0,3)$-form or BF theories with suitable interactions on twistor space. This partially reveals their quasi-topological nature and classical integrability. In this section, we study their BV-BRST formulation as a preparation for the study of quantization at one loop.

\subsubsection{BV-BRST formalism for holomorphic Chern-Simons theories}
\paragraph{BV parent actions.} To quantize the holomorphic Chern-Simons theories in a neat way, we will employ the BV-BRST formalism adapted to the holomorphic setting, cf. \cite{williams2020renormalization,Costello:2021bah,Bittleston:2022nfr}. Namely, we will consider the BV fields\footnote{See e.g. \cite{Cattaneo:2019jpn} for a beautiful introduction to BV-BRST formalism. However, we will need a bit more than the usual BV formalism. In particular, we will need a regularized version of BV formalism where the BV theory is defined within a range of length scale. See below and also Appendix \ref{app:RGEFT} for a short review.}
\begin{align}
    \A_{2h-2}\in \Omega^{0,\bullet}(\PT,\cO(2h-2)\otimes\mg)[1]\,,
\end{align}
with all anti-holomorphic (Dolbeault) form degrees saturated. Here, $[1]$ denotes a shift in cohomological degree or ghost degree. In particular, $f\in \Omega^{0,k}(\PT)[1]$ will be assigned cohomological degree or ghost degree $1-k$. %
With this consideration, the BV field $\A$ decomposes as
\begin{align}\label{eq:BV-field}
    \A_{2h-2}=\cc_{2h-2}+\sA_{2h-2}+\sA^{\vee}_{2h-2}+\cc^{\vee}_{2h-2}\,,
\end{align}
where the field components are\footnote{We consider $\mg$ to be semisimple Lie algebra, so that the dual Lie algebra $\mg^{\vee}$ is isomorphic with $\mg$.}
\begin{subequations}
    \begin{align}
     \text{ghost}&:   &\cc_{2h-2}&\in \Omega^{0,0}(\PT,\cO(2h-2)\otimes\mg)[1]\,, &\gh(\cc)&=1\,,\\
      \text{field}&:  &\sA_{2h-2}&\in \Omega^{0,1}(\PT,\cO(2h-2)\otimes \mg)[1]\,, &\gh(\sA)&=0\,,\\
      \text{antifield}&:  &\sA^{\vee}_{2h-2}&\in \Omega^{0,2}(\PT,\cO(2h-2)\otimes \mg)[1]\,,&\gh(\sA^{\vee})&=-1\,,\\
       \text{antifield of ghost}&: &\cc^{\vee}_{2h-2}&\in \Omega^{0,3}(\PT,\cO(2h-2)\otimes\mg)[1]\,,&\gh(\cc^{\vee})&=-2\,.
    \end{align}
\end{subequations}
The BV actions associated with the BV fields \eqref{eq:BV-field} read
\small
\begin{subequations}\label{eq:BV-action}
    \begin{align}
    S_{BV-hCS}^{\star}&=\int_{\PTc}\Omega^{3,0}\Tr\Big(\sum_h\A_{-2|h|-2}\bar{\p}\A_{2|h|-2}+\frac{2}{3}\sum_{\{h_i\}}\frac{1}{k!}\A_{2h_1-2}\Pi^k(\A_{2h_2-2},\A_{2h_3-2})\Big)\,,\\
    S_{BV-hCS}^{\{,\}}&=\int_{\PTc}\Omega^{3,0}\Big(\sum_h\A_{-2|h|-2}\bar{\p}\A_{2|h|-2}+\frac{1}{3}\sum_{\{h_i\}}\A_{2h_1-2}\{\A_{2h_2-2},\A_{2h_3-2}\}\Big)\,,\\
    S_{BV-hCS}^{[,]_{\mg}}&=\int_{\PT}\Omega^{3,0}\Tr\Big(\sum_h\A_{-2|h|-2}\bar{\p}\A_{2|h|-2}+\frac{1}{3}\sum_{\{h_i\}}\A_{2h_1-2}[\A_{2h_2-2},\A_{2h_3-2}]\Big)\,,
\end{align}
\end{subequations}
\normalsize
where for the case of $S_{BV}^{\star}$, $k$ obeys the same helicity constraint as in \eqref{eq:helicity-constraint-star}. 

As usual, the space of BV fields $\cF_{BV}$ is naturally a differential graded symplectic manifold endowed with a degree $-1$ symplectic form $\omega$. This symplectic form induces the BV holomorphic symplectic pairing 
\begin{align}
    \langle a,b\rangle_{\PTc}:=\int_{\PTc} \Omega^{3,0}
    \Tr(a\,b)\,,
\end{align}
on the space of BV fields $\cF_{BV}$,
where $a\in \Omega^{0,p}(\PT, \mg)[1]\,,\ b\in\Omega^{0,3-p}(\PT,\mg)[1]$. 

Using this geometric data, we can then read off the cohomological vector fields $Q$ associated to the actions \eqref{eq:BV-action} as 
\begin{subequations}\label{eq:BRST-transform}
    \begin{align}
        Q_{\star}&=\int_{\PTc}\Omega^{3,0}\Tr\Big(\F_{\star}\frac{\delta}{\delta \A}\Big)\,, 
        \qquad &\F_{\star}&=\bar{\p}\A+\frac{1}{2}[\A,\A]_{\star}\,,\\
        Q_{\{,\}}&=\int_{\PTc}\Omega^{3,0}\Big(\F_{\{,\}}\frac{\delta}{\delta \A}\Big)\,, 
        \qquad &\F_{\{,\}}&=\bar{\p}\A+\frac{1}{2}\{\A,\A\}\,,\\
        Q_{[,]}&=\int_{\PT}\Omega^{3,0}\Tr\Big(\F_{[,]}\frac{\delta}{\delta \A}\Big)\,, 
        \qquad &\F_{[,]}&=\bar{\p}\A+\frac{1}{2}[\A,\A]\,,
    \end{align}
\end{subequations}
where we will temporarily suppress the helicity labels for a neat discussion. 
As usual, the equations of motion for the field components can be obtained from
\begin{align}
    Q_{\bullet} S_{BV-hCS}^{\bullet}=0\,,\qquad \bullet=\star,\ \{\, ,\},\ [\,,]\,.
\end{align}
Then, upon restricting to appropriate sectors, we can read off the BRST transformations for the physical field $\sA\in \Omega^{0,1}(\PT,\cO(n)\otimes\mg)[1]$ and the ghost $\cc\in \Omega^{0,0}(\PT,\cO(n)\otimes\mg)[1]$, for instance, as
\begin{subequations}
    \begin{align}
    \delta \sA&=Q_{\star}\,\sA\Big|_{(0,1)\text{-part}}=\bar{\p}\cc+[\sA,\cc]_{\star}\,,\\
    \delta \cc&=Q_{\star}\,\cc\Big|_{(0,0)\text{-part}}=\frac{1}{2}[\cc,\cc]_{\star}\,.
\end{align}
\end{subequations}
Other BRST gauge transformations can also be obtained for the cases where the BV actions are $S^{\{,\}}_{BV}$ and $S^{[,]_{\mg}}_{BV}$, so we leave them as an exercise.

\paragraph{Gauge fixing.} As usual, to quantize the above actions, we need to choose a gauge and reduce the BV phase space to a Lagrangian isotropic subspace $\cL_{\Psi}\subset \cF_{BV}$. 
This can be done by considering a gauge-fixing fermion function of cohomological degree $-1$
\begin{align}
    \Psi=\Psi(\bar{\cc},\sA,\sn)\,,
\end{align}
which is, in general, a function of antighost $\bar{\cc}$, the physical field $\sA$, and the Nakanishi-Lautrup field $\sn$. It is useful recall that $\bar{\cc}$ and $\sn$ are not part of the original BV fields we start with. 
They only arise when we gauge fix the BV action to obtain the partition function
\begin{align}
    \cZ=\int_{\cL_{\Psi}}D\A \exp\Big(S^{gf}_{BV-hCS}\Big[\sA,\cc,\bar{\cc},\sn,\sA^{\vee}=\frac{\delta \Psi}{\delta \sA},\bar{\cc}^{\vee}=\frac{\delta \Psi}{\delta \bar{c}}\Big]\Big)
\end{align}
where now
\begin{align}
    \A=\big(\sA,\cc,\bar{\cc},\sn\big)\,.
\end{align}
Note that the restriction to a Lagrangian submanifold is the statement of eliminating all antifields and antifields of antighosts 
by the substitution
\begin{align}
    \sA^{\vee}=\frac{\delta\Psi}{\delta\sA}\,,\qquad \bar{\cc}^{\vee}=\frac{\delta \Psi}{\delta \bar{\cc}}\,.
\end{align}
This leaves us with a BV action that contains only the physical fields $\sA$, the ghosts $\cc$, the anti-ghosts $\bar{\cc}$, and the Lagrangian multiplier $\sn$ with ghost degree 0.  

For holomorphic Chern-Simons and BF-type theories considered in this work, we take
\begin{align}\label{eq:GF-hCS}
    \Psi=\int_{S^7}d\theta \Omega^{3,0}\Big(\bar{\cc}\,\bar{\p}^{\dagger}\sA\Big)\,.
\end{align}
where
\begin{align}
    \bar{\cc}=\sum_{h}e^{i\theta(2h-2)}\bar{\cc}_{2h-2}\,,\qquad \bar{\cc}_{2h-2}\in \Omega^{0,3}(\PT,\cO(2h-2)\otimes\mg)[2]\,,\quad \gh(\bar{\cc})=-1\,,
\end{align}
as the minimal gauge-fixing fermion. That is, we impose the gauge fixing conditions
\begin{align}\label{eq:gauge-fixing}
    \bar \p^{\dagger}\sA=0\,,\qquad  \bar{\p}^{\dagger}=-\ast \bar\p\,\ast:\Omega^{0,k}(\PT)\rightarrow\Omega^{0,k-1}(\PT)\,,
\end{align}
where $\ast:\Omega^{p,q}\rightarrow \Omega^{3-p,3-q}$ is 
the usual Hodge operator associated with the Hermitian metric on $\PT$. As discussed, we can now solve for the antifield of $\bar{\cc}$, which is
\begin{align}
    \bar{\cc}^{\vee}=\sum_{h}e^{i\theta(2h-2)}\bar{\cc}^{\vee}_{2h-2}\,,\qquad \bar{\cc}^{\vee}_{2h-2}\in \Omega^{0,0}(\PT,\cO(2h-2)\otimes\mg)\,,\quad \gh(\bar{\cc}^{\vee})=0\,,
\end{align}
to project our partition function to the Lagrangian submanifold $\cL_{\Psi}\subset \cF_{BV}$. The gauge-fixed holomorphic Chern-Simons BV action reads 
\small
\begin{align}
    S^{gf}_{BV-hCS}=\int_{\PTc}\Omega^{3,0}\Tr\Bigg(\sum_h\Big(\sA_{-2|h|-2}\bar{\p}\sA_{2|h|-2}+\bar{\cc}_{-2h-2}\bar{\p}^{\dagger}\bar{\p}\cc_{2h-2}+\sn_{-2h-2}\bar{\p}^{\dagger}\sA_{2h-2}\Big)+\cV(\A^3)\Bigg)\,,
\end{align}
\normalsize
where $\sn\in\Omega^{0,3}(\PT,\mg)$ with ghost number 0 is the Nakanishi-Lautrup aka Langragian multiplier, and the $\cV(\A^3)$ denotes the expansion of the cubic vertices in terms of the field components of $\A$. 
\subsubsection{BV-BRST formalism for holomorphic BF theories}
\paragraph{BV parent actions.} The BV-BRST formulation of BF and BF-Poisson theories, which are the twistor duals of self-dual YM and self-dual gravity, has been studied in \cite{Bittleston:2022nfr}. Here, we only attempt to extend their results to the higher-spin case. 

The BV field content for holomorphic BF-type theories are slightly different with hCS. In particular, we need to consider two set of BV fields
\begin{subequations}\label{eq:BF-BV}
    \begin{align}
    \A_{2s-2}&\in\Omega^{0,\bullet}(\PT,\cO(2s-2)\otimes\mg)[1]\,,\\\B_{-2s-2}&\in\Omega^{0,\bullet}(\PT,\cO(-2s-2)\otimes\mg)[1]\,,
\end{align}
\end{subequations}
where their field components are
\begin{subequations}
    \begin{align}
        \A_{2s-2}&=\cc_{2s-2}+\cA_{2s-2}+\cB^{\vee}_{2s-2}+\dd^{\vee}_{2s-2}\,,\\
        \B_{-2s-2}&=\dd_{-2s-2}+\cB_{-2s-2}+\cA^{\vee}_{-2s-2}+\cc^{\vee}_{-2s-2}\,.
    \end{align}
\end{subequations}
The notions of (fields, ghosts) versus  (antifields, antifield of ghosts) should be clear from the context. Here, the ghost $\cc$ can be identified with the gauge parameter $\xi$, and the ghost $\dd$ can be identified with $\chi$ in \eqref{eq:gauge-variation-BF}. For simplicity, we can assume that $\mg$ is chosen such that $\mg^{\vee}=\mg$. 
The BV actions for BF theories read
\begin{subequations}\label{eq:BV-action-BF}
    \begin{align}
    S_{BV-BF}^{\{,\}}&=\int_{\PTc}\Omega^{3,0}\Big(\sum_s\B_{-2s-2}\bar{\p}\A_{2s-2}+\frac{1}{2}\sum_{\{s_i\}}\B_{-2s_1-2}\{\A_{2s_2-2},\A_{2s_3-2}\}\Big)\,,\\
    S_{BV-BF}^{[,]_{\mg}}&=\int_{\PT}\Omega^{3,0}\Tr\Big(\sum_s\B_{-2s-2}\bar{\p}\A_{2s-2}+\frac{1}{2}\sum_{\{s_i\}}\B_{-2s_1-2}[\A_{2s_2-2},\A_{2s_3-2}]\Big)\,,
\end{align}
\end{subequations}
The cohomological vector fields associated to the above action are
\begin{subequations}
    \begin{align}
        Q_{\{,\}}&=\int_{\PTc}\Omega^{3,0}\Big(\F_{\{,\}}\frac{\delta}{\delta \B}+\bar{\cD}_{\A}\B\frac{\delta}{\delta \A}\Big)\,, 
         &\bar{\cD}_{\A}&=\bar{\p}+\{\A,-\}\,,\\
        Q_{[,]}&=\int_{\PT}\Omega^{3,0}\Tr\Big(\F_{[,]}\frac{\delta}{\delta \B}+\bar{D}_{\A}\B\frac{\delta}{\delta \A}\Big)\,, 
         &\bar{D}_{\A}&=\bar{\p}+[\A,-]\,,
    \end{align}
\end{subequations}
where the curvature $\F_{\bullet}$ are defined similar to the ones in \eqref{eq:BRST-transform}, but with $\A$ containing only fields of positive helicities. The gauge fixing procedure for \eqref{eq:BV-action-BF} can be done anologously with the case of hCS. We refer the reader to \cite{Bittleston:2022nfr} for further detail.
\section{Quantization of holomorphic higher-spin theories at one loop}\label{sec:4}
Armed with the BV action \eqref{eq:BV-action} for all holomorphic higher-spin theories on twistor space, we now turn to its quantization at one loop. In particular, we shall resort to e.g. \cite{williams2020renormalization, Costello:2021bah, Bittleston:2022nfr} for the recent advancements in computing the anomaly for holomorphic theories. What we find is that there are certain holomorphic higher-spin theories that are self-quantum consistent, meaning they are anomaly-free at one loop without requiring any counterterms. Notably, one can determine whether a theory is anomaly-free based on its spectrum, a result stemming from the severe constraints imposed by the higher-spin symmetry on all models considered in this work. For anomalous theories, we extend the anomaly cancellation mechanism introduced in \cite{Costello:2021bah, Bittleston:2022nfr} to the case of higher spin to render anomalous twistorial higher-spin theories quantum consistent at one loop. 

\subsection{A higher-spin index theorem}\label{sec:index}
This subsection derives an index theorem for holomorphic twistorial higher-spin theories, which is a higher-spin generalization of the well-known Hirzebruch-Riemann-Roch index theorem. Note that our index can be associated to the integrable background complex structure $\bar{\p}$ twisted by a holomorphic vector bundle $\V$ (denoted as $\bar{\p}_\V$) if we consider flat twistor space, or $\bar{\cD}=\bar{\p}+\Pi(\sA_2,-)$ if we consider curved twistor space. As in the usual context, the index will tell us whether a twistorial higher-spin theory may possess gauge, gravitational or mixed type anomalies at one loop. 

Recall that the kinetic action for the parent BV action, cf. \eqref{eq:BV-action}, when coupled to a gravitational background is\footnote{Here, we shall proceed with holomorphic Chern-Simons theories, as similar computations can be carried out for the case of holomorphic BF theories.}
\begin{align}\label{eq:kin-action}
    S^{kin}_{BV-hCS}&=\int_{\PTc}\Omega^{3,0}\Tr\Big(\sum_h\A_{-2|h|-2}\bar{\p}_{\V}\A_{2|h|-2}+\sum_h\A_{-2|h|-2}\Pi(\sA_2,\A_{2|h|-2})\Big)\nn\\
    &\equiv\int_{\PTc}\Omega^{3,0}\Tr\Big(\sum_h\A_{-2|h|-2}\bar{\cD}_{\V}\,\A_{2|h|-2}\Big)\,.
\end{align}
As mentioned briefly above, 
\begin{align}\label{eq:SD-connection-twistor}
    \bar{\cD}_{\V}:=\bar{\p}_{\V}+\Pi(\sA_2,-)\,,\qquad \sA_2\in \Omega^{0,1}(\PT,\cO(2))[1]\,.
\end{align}
is referred to as the deformed complex structure sourced by the Hamiltonian flow $\Pi(\sA_2,-)$ and twisted by a holomorphic vector bundle $\V$.

Note that the connection $\bar{\cD}_{\V}$ can be shown to be in one-to-one correspondence with the SD connection of the corresponding SD spacetime through the Penrose non-linear graviton construction \cite{Penrose:1976js}. The Hirzebruch-Riemann-Roch theorem  states that the index associated to the operator $\bar{\cD}_{\V}$ is given by
\begin{align}
    \Ind (\bar{\cD}_{\V})=\int_{\PT}\Tod(T_{\PTc})\wedge\Ch(\V)\,.
\end{align}
Here, $\Tod(T_{\PTc})$ is the Todd class associated with the tangent bundle $T_{\PTc}$. Furthermore, $\Ch(\V)$ stands for the Chern character of the vector bundle $\V$ where the BV fields $\A$ take values in. The vector bundle $\V$ is given by:
\begin{align}
    \V:=\mg\otimes \bigoplus_{h\in \Spec}\cO(2h-2)\,,
\end{align}
where $\Spec$ is the spectrum of fields of the theory in consideration.\footnote{A critical reader may be slightly annoyed by the way we abuse the terminology, since the index should only be associated to the zero modes or ground states of the fields encoded by $\Spec$ following the usual Fredholm index formula $\Ind(\bar{\cD}_{\V})=\dim\ker \bar{\cD}_{\V}-\dim\mathrm{coker} \bar{\cD}_{\V}=\dim \ker \bar{\cD}_{\V}-\dim \bar{\cD}_{\V}^{\dagger}$.}

Let us now examine the detailed elements of the above index $\Ind(\bar{\cD})$. First, the Todd class associated with $T_{\PTc}$ reads
\begin{align}
    \Tod(T_{\PTc})=1+\Tod_1(T_{\PTc})+\Tod_2(T_{\PTc})+\Tod_3(T_{\PTc})+\Tod_4(T_{\PTc})+\ldots
\end{align}
where
\begin{align}
    \Tod_1(T_\PTc)=\frac{\tc_1(T_{\PTc})}{2}\,,\qquad \Tod_2(T_\PTc)=\frac{\tc_1^2(T_{\PTc})+\tc_2(T_{\PTc})}{12}\,\quad \text{etc.,}
\end{align}
\normalsize
with $\tc_{n\geq 1}(T_{\PTc})$ being the Chern classes of $T_\PTc$. (A review of characteristic classes can be found in e.g. \cite{milnor1974characteristic,Nakahara:2003nw}.) 

To proceed we recall that for $V_i$ being some vector bundles and $\sF$ their curvature 2-form, we have
\small
\begin{subequations}
    \begin{align}
        \Ch(V_1\otimes V_2)&=\Ch(V_1)\wedge \Ch(V_2)\,,\\
        \Ch(V_1\oplus V_2)&=\Ch(V_1)+\Ch(V_2)\,,\\
        \Ch(V)&=\Tr\, \exp\Big[-\sum_k\frac{1}{k}\Big(\frac{\sF}{2\pi i}\Big)^k\Big]=\rank(V)+\Ch_1(\sF)+\Ch_2(\sF)+\Ch_3(\sF)+\ldots\,,
    \end{align}
\end{subequations}
\normalsize
where 
\begin{align}
    \Ch_1&=\tc_1\,,\quad \Ch_2=\frac{1}{2}\Big[\tc_1^2-2\tc_2\Big]\,,\quad \Ch_3=\frac{1}{3!}\Big(\tc_1^3+3\tc_1\tc_2-3\tc_3\Big)\,,
    \quad \text{etc.}
\end{align}
\normalsize
Here, $\cc_n\equiv \frac{i^n}{n!(2\pi)^n}\Tr(\sF^n)$ denotes the $n$th Chern class.  

With the above information, we can reduce the computation of the index $\Ind(\bar{\cD}_{\V})$ to\footnote{Have we considered the index $\Ind(\bar{\p}_{\V})$, the corresponding formula would  be almost the same, with the substitution $T_{\PTc} \mapsto T_{\PT}$.} 
\begin{align}\label{eq:pre-HS-index}
    \Ind (\bar{\cD}_{\V})=\int_{\PTc}\Tod(T_{\PTc})\wedge \Ch(\mg)\wedge  \Bigg[\sum_{h\in \Spec}\Ch\Big(\cO(2h-2)\Big)\Bigg]_{(1,1)}\,.
\end{align}
It turns out that there is one more simplification. Namely, the Chern character of $\cO(n)$ gets truncated at the first Chern class \cite{Bittleston:2022nfr}, i.e. 
\begin{align}
    \Ch\Big(\cO(n)\Big)_{(1,1)}=1+\tc_1(\cO(n))=1-\frac{n}{2}\tK\,,
\end{align}
where 
\begin{align}
    \tK=\frac{\langle \lambda d\lambda\rangle\wedge\langle \hat \lambda d\hat \lambda\rangle}{\langle \lambda\,\hat\lambda\rangle^2}
\end{align}
is the volume form of $\P^1$. 
\begin{theorem} All higher-spin theories that arise from the action \eqref{eq:BV-action} on twistor space with trivial index $\Ind(\bar{\cD}_{\V})$ are anomaly-free at one loop.
\end{theorem}

The above theorem can be proven through the computation in the next subsection. In particular, the trivialization of the wheel amplitude, cf. Fig \ref{fig:anomaly}, is closely related to the spectrum of the theory under consideration. Due to the fact that the spectrum of higher-spin theories are typically $\infty$-dimensional, the regularization scheme that we will use for sum over the spectrum deserves a paragraph for explanation.

As it is well-known, there are some higher-spin theories in AdS that can be defined holographically. In particular, Klebanov and Polyakov \cite{Klebanov:2002ja} conjectured that higher-spin gravities in $AdS_{d+1}$ should be dual to free or critical vector models on the boundary of AdS, see also \cite{Sezgin:2002rt}. Since then, there have been many non-trivial tests for matching the one-loop vacuum energies on both sides of the duality (see e.g. \cite{Giombi:2013fka,Giombi:2014yra}). In particular, for gravitational theories in AdS whose spectrum contain spin $s=0,1,\ldots,\infty$, the one-loop determinants in AdS typically trivialize under Riemann zeta regularization \cite{Beccaria:2015vaa}, which are in complete agreement with the dual CFT results. However, for theories with only even spins, the one-loop determinants can result in non-trivial and interesting numbers, which again can be checked to match precisely with the CFT's computations, see e.g. \cite{Giombi:2014xxa}. This justifies the use of Riemann zeta regularization to handle the infinite sums over the spectrum in higher-spin theories. And, in fact, this choice of regularization is natural, since higher-spin theories are expected to be certain limits of string theories, see e.g. discussion in \cite{Engquist:2005yt}.

Now, coming back to our problem, and consider the case where $\Spec=\Z$. We get
\begin{align}
    \Bigg[\sum_{h\in \Z}\Ch\Big(\cO(2h-2)\Big)\Bigg]_{(1,1)}=(1+\tK)+2\sum_{n\geq 1}^{\infty}(1+\tK)=(1+\tK)\Big(1+2\zeta(0)\Big)=0\,.
\end{align}
Therefore, 
\begin{align}\label{eq:HS-index}
    \Ind(\bar{\cD}_{\V})=0\quad \text{for}\quad \Spec=\Z\,.
\end{align}
Observe that $\Tod(T_{\PTc})$ and $\Ch(\mg)$ do not play crucial roles in trivializing $\Ind(\bar{\cD}_{\V})$ since they form an overall factor in the integrand. Remarkably, this type of sum also appears recently in the study of one-loop partition function of M2-brane in M-theory on $AdS_4\times S^7/\Z_k$ background, cf. \cite{Beccaria:2023ujc}. 

Although the above theorem is essentially true for higher-spin theories with $\Spec=\Z$ or $\Spec=2\Z^*$, we can also consider theories with spectrum of only odd spins, where it can be shown that\footnote{To regularize the sum, we use $\zeta(s)=\sum_{n=1}^{\infty}\frac{1}{n^s}$ and take $s$ to zero limit at the end.} 
    \small
    \begin{align}
        \Bigg[\sum_{h\in 2\Z+1}\Ch\Big(\cO(2h-2)\Big)\Bigg]_{(1,1)}=2(1+\tK)\sum_{\substack{n=2k+1\\
        k\geq 0}}^{\infty}1=(1+\tK)\Bigg(\lim_{s\rightarrow 0}\sum_{n\geq 1}\frac{1}{(2n-1)^s}\Bigg)=0\,,
    \end{align}
    \normalsize
    which again results in a trivial index. However, in this case, the corresponding self-dual spacetime background is generated by a positive-helicity spin-1 field, and as a result, $\PT$ remains topologically trivial, see e.g. \cite{Ward:1977ta,Adamo:2024xpc}. 
    
    Bearing in mind that theories with $\Spec=\{|h|\geq 2\}$ has non-trivial contributions to the index $\Ind(\bar{\cD}_{\V})$ since 
    \begin{align}
        \sum_{|h|\geq 2}\Ch\big(\cO(2h-2)\big)=2(1+\tK)\sum_{n=2}^{\infty}1=-3(1+\tK)\neq 0\,.
    \end{align}
    Nevertheless, this can be remedied away by introducing a scalar field and a spin-1 gauge field into the spectrum enhancing $\Spec=\{|h|\geq 2\}$ to $\Spec=\Z$.

    To this end, it is useful to note that the index above does not specify the exact form of the anomalies, which obstruct quantum consistency of many higher-spin theories derived from \eqref{eq:BV-action}. In particular, one needs additional details to compute the anomalies exactly, especially in the case of higher-derivative interactions. This will be done in the next subsection.

    \paragraph{Stringy features of holomorphic higher-spin theories.} Let us return to the formula \eqref{eq:pre-HS-index} and expand the Todd class and Chern character up to the 4th order. Using the fact that all odd-th Chern classes $\tc_n(\mg)$ of semisimple Lie algebra $\mg$ vanish, i.e. $\tc_{2j+1}(\mg)=0$, and following the steps in \cite{Bittleston:2022nfr}, we obtain 
    \begin{align}
        \Ind(\bar{\D}_{\V})=\Ind(\bar{\cD}_{\V})_{gauge}+\Ind(\bar{\cD}_{\V})_{grav}+\Ind(\bar{\cD}_{\V})_{mixed}\,,
    \end{align}
    where
    \begin{subequations}
        \begin{align}
            \Ind(\bar{D}_{\V})_{gauge}&=\int_{\PTc}\Ch_4(\mg)\wedge  \Bigg[\sum_{h\in \Z}\Ch\Big(\cO(2h-2)\Big)\Bigg]_{(1,1)}\,,\\
            \Ind(\bar{D}_{\V})_{grav}&=\frac{1+\dim(\mg)}{20}\int_{\PTc}\Ch_4(T_{\PTc})\wedge  \Bigg[\sum_{h\in \Z}\Ch\Big(\cO(2h-2)\Big)\Bigg]_{(1,1)}\,,\\
            \Ind(\bar{D}_{\V})_{mixed}&=-\frac{1}{12}\int_{\PTc}\Ch_2(T_{\PTc})\wedge\Ch_2(\mg)\wedge  \Bigg[\sum_{h\in \Z}\Ch\Big(\cO(2h-2)\Big)\Bigg]_{(1,1)}\,.
        \end{align}
    \end{subequations}
    In the above, $\Ind(\bar{D}_{\V})_{gauge}$, $\Ind(\bar{D}_{\V})_{grav}$, and $\Ind(\bar{D}_{\V})_{mixed}$ may be referred to as the indices associated with gauge, gravitational, and mixed-type anomalies. 
    
    Since all holomorphic higher-spin theories derived from the BV action \eqref{eq:BV-action}, have a spectrum of infinitely many massless fields and exhibit some topological nature, one may expect that they can be referred to as the field theory descriptions of certain higher-spin generalization of the topological string theories, cf. \cite{Witten:1988xj,Witten:2003nn,Alexandrov:1995kv}. 
    Thus, ultimately, these theories should be interpreted as some sort of string theories. 
    We find this perspective intriguing since it is partially supported by the recent study on coupling chiral/self-dual higher-spin theories (the spacetime duals of holomorphic twistorial theories) with a test particle in flat space, cf. \cite{Ivanovskiy:2025kok}. There, it is observed that chiral higher-spin gravity cannot couple to a test particle, suggesting that this theory may couple to extended objects. This is expected, as only extended objects can accommodate all the degrees of freedom of higher-spin theories. 

    


\subsection{Anomaly-free twistorial theories}\label{sec:loop-computation}
Having established that theories with a trivial index associated with $\bar{\cD}$ or $\bar{\p}$ should be anomaly-free at one loop, we now proceed to show in this subsection that this is indeed the case.

\paragraph{The anomaly.} It is useful recalling that the $n$-th Chern character can be written in curvature form as
\begin{align}
    \Ch_{n}(\sF)=\frac{i^{n}}{n!(2\pi)^n}\Tr\big(\sF^{\wedge n}\big)\,.
\end{align}
The gauge anomaly for a non-abelian gauge theory in a $2n$-dimensional real space $\cM$ can be obtained directly from the anomaly polynomial and has the form
\begin{align}
    \mathscr{C}=\frac{i^{n}}{n!(2\pi)^n}\int_{\cM}\Tr\Big(\cc \wedge \sF^{\wedge n}\Big) \,,
\end{align}
where $\cc$ is the ghost (or gauge parameter), and $\sF$ is the corresponding field strength of the theory. Note that in $4n+2$ real dimensions, we also have gravitational anomaly, which happens to be the case here since $\dim_{\R}(\PTc)=6$. For standard treatment of anomaly, see e.g. \cite{Alvarez-Gaume:1984zlq,Bilal:2008qx,Fujikawa:2004cx}.

\begin{figure}[h!]
    \centering
    \includegraphics[scale=0.3]{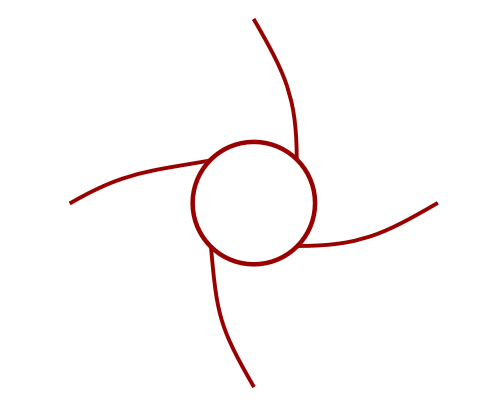}
    \caption{The Feynman wheel diagram associated with the gauge anomaly in $6d$.}
    \label{fig:anomaly}
\end{figure}

In what follows, we will not display the gravitational anomaly in terms of the curvature 2-form $\mathsf{R}$ since we observe that there is a uniform way of writing the anomalies for various holomorphic twistor theories with higher-derivative interactions (see below). To briefly summarize our result, it is convenient to introduce the operator
\begin{align}
    \mho_{\star}:=\frac{\Big([\p_2\,\p_3]+[\p_2\,\p_4]+[\p_3\,\p_4]\Big)^{\tH_4-4}}{(\tH_4-4)!}\,,\qquad \p_{i}=\frac{\p}{\p\mu_i^{\dot\alpha}}\,,
\end{align}
where $\tH_4=h_1+h_2+h_3+h_4$ with $h_i$ are the helicities of the external fields entering the wheel diagram Fig. \ref{fig:anomaly}. 

Then, the most general anomaly for holomorphic twistorial theories derived from \eqref{eq:BV-action} with $\Spec=\Z$ 
is 
\begin{align}\label{eq:general-anomaly}
    \Ccurl_{hBV^{\star}}=\sum_{\Spec=\Z}\frac{i^3}{3!(2\pi)^3}\int_{S^7\simeq U(1)\times \PTc} d\theta\,\mho_{\star}\,\Tr\Big(\A\wedge \F_2 \wedge\F_3\wedge \F_4\Big)\Bigg|_{\substack{\F_{2,3,4}=\F\\
    \p_{2,3,4}\mapsto\p}}\,,
\end{align}
where the field strength reads
\begin{align}
    \F_i=d\A_i +\A_i\star\A_i\,,
\end{align}
and
\begin{align}
    \A_i=\sum_he^{i\theta(2h-2)}\A_{i,2h-2}\,,\qquad \A_{i,n}\in \Omega^{0,\bullet}(\PT,\cO(n)\otimes\mg)[1]\,,
\end{align}
are the associated higher-spin fields. As the notation suggested, it is understood that, in the above expression of $\Ccurl_{hBV}^{\star}$, all derivatives are evaluated at some point $z$ on $\PTc$, cf. \eqref{eq:general-anomaly}. Once again, we use the monodromy trick and work with $S^7\simeq U(1)\times \PTc$, and adopt BV notation. 

Decomposing the de Rham differential as 
\begin{align}
    d=\p+\bar{\p}=dZ^A\frac{\p}{\p Z^A}+d\hat{Z}^A\frac{\p}{\p \hat Z^{A}}\,,\qquad A=1,2,3,4\,,
\end{align}
we can compute the anomaly $\Ccurl_{hBV^{\star}}$, which yields
\begin{align}
    \mathscr{C}_{hBV^{\star}}=\sum_{\Spec=\Z}\frac{i^3}{3!(2\pi)^3}\int_0^{2\pi} d\theta \int_{\PTc}\mho_{\star}\Tr\Big(\A\wedge (\p \A_2)\wedge (\p\A_3)\wedge(\p\A_4)\Big)\Bigg|_{\substack{\A_{2,3,4}=\A\\
    \p_{2,3,4}\mapsto\p}}
    \,,
\end{align}
where it is useful noting that contributions associated with the non-linear terms in the field strength are suppressed by virtue of form degrees. Upon restricting the BV fields to their physical and the ghost components, we obtain 
\begin{align}\label{eq:anomaly-star}
    \mathscr{C}_{hBV^{\star}}=\sum_{\Spec=\Z}\frac{i^3}{3!(2\pi)^3}\int_0^{2\pi} d\theta \int_{\PTc}\mho_{\star}\Tr\Big(\cc\wedge (\p \sA_2)\wedge(\p\sA_3)\wedge(\p\sA_4)\Big)\Bigg|_{\substack{\sA_{2,3,4}=\sA\\
    \p_{2,3,4}\mapsto\p}}
    \,.
\end{align}
From here, our direct computation for the above wheel diagram will be proceeded along the lines of \cite{williams2020renormalization} and \cite{Bittleston:2022nfr}, with some modifications to account for the interaction involving the $\star$-product. Note that truncations to $[\,,]_{\mg}$ and $\{\,,\}$ cases can be done with no ambiguity.

\paragraph{Computing the wheel diagram.} Since quantum anomaly is a local effect, we can consider the patch $\C^3\subset \PTc$ where $\{\lambda\neq u\}$ for $u$ a constant spinor, cf. \cite{williams2020renormalization,Bittleston:2022nfr}. On this patch, we will use the following in-homogeneous coordinates
\begin{align}
    z^{0}=\frac{\langle \lambda \,u\rangle}{\langle \lambda\,\hat u\rangle}\,,\qquad z^{\dot\alpha}=\frac{\mu^{\dot\alpha}}{\langle\lambda\,u\rangle}\,,\qquad \langle u\,\hat u \rangle=1\,.
\end{align}

Observe that we can trivialize $\cO(n)$ by some function $F(\lambda,u)=f/\langle \lambda\,\hat u\rangle^n$. With this, the holomorphic measure on $\PTc$ can be written as
\begin{align}
     \frac{\Omega^{3,0}}{\langle \lambda\,\hat u\rangle^4}=\frac{1}{\langle \lambda\,\hat  u\rangle^4}\langle \lambda\,d\lambda\rangle \wedge [d\mu\wedge d\mu]= dz^1dz^{\dot 1}dz^{\dot 2}\equiv dz^1dz^2dz^3\,.
\end{align}
As usual, to compute observables of a theory, one needs to choose a gauge. For actions formulated in first-order formalism, a convenient choice is $\bar{\p}^{\dagger}\sA=0$, cf. \eqref{eq:gauge-fixing}. In this gauge, 
\begin{align}
    \bar{\p}^{\dagger}P(z_1,z_2|\ell_1,\ell_2)=0\,,
\end{align}
where $P(z_1,z_2|\ell_1,\ell_2)$ is a regularized propagator between two points $z_1,z_2$ in $\C^3$; and $\ell_1\leq \ell_2$ are some characteristic length scales introduced to describe the behavior of the propagators within the range $[\ell_1,\ell_2]$. Henceforth, we will interpret $\ell_1\mapsto \varepsilon$ as a UV cutoff, and $\ell_2\mapsto L$ as an IR cutoff.\footnote{For a more detailed discussion of the roles of $\ell_1,\ell_2$, see Appendix \ref{app:RGEFT} and references therein.}

In the above, $P(z_1,z_2|\ell_1,\ell_2)$ is 
a matrix-valued $(0,2)$-form on $\C^3\subset \PTc$ obeying
\begin{align}\label{eq:bar-p-P}
    \bar{\p}P^{h_1,h_2}(z_1,z_2|\ell_1,\ell_2)=-\delta_{h_1+h_2,0}\Big(\bigwedge_{j=1}^3(d \bar{z}_1^j-d\bar{z}_2^j)\Big)\Big[k_{\ell_1}(z_{12})-k_{\ell_2}(z_{12})\Big]\,,
\end{align}
where the Kronecker delta $\delta_{h_1+h_2,0}$ indicates the exchange in helicity, and
\begin{align}
   k_{\ell}(z_{12}):=\Big(\frac{1}{4\pi\ell}\Big)^3e^{-\frac{|z_{12}|^2}{4\ell}}\,,
\end{align}
denotes the corresponding heat kernel. 

Since anomaly is an UV effect, we can safely take the limit $L \rightarrow \infty$ at this stage, while keeping $\varepsilon$ to be small. With this consideration
\begin{align}
    P^{h_1,h_2}(z_1,z_2|\varepsilon,\infty)=-\delta_{h_1+h_2,0}\,\Omega_{12}^{(0,2)}\int_{\varepsilon}^{\infty}\frac{d\ell}{2\ell} \Big(\frac{1}{4\pi\ell}\Big)^3 e^{-\frac{|z_{12}|^2}{4\ell}}\,
\end{align}
where
\begin{align}
    \Omega^{(0,2)}_{12}=\epsilon_{abc}\bar{z}_{12}^ad\bar{z}_{12}^bd\bar{z}_{12}^c\,,\qquad z_{12}=z_1-z_2\,,\qquad a,b=1,2,3\,.
\end{align}
To concisely streamline our analysis, we will denote
\begin{subequations}
\begin{align}
    P[h_i,h_j]&\equiv P^{2h_i-2,2h_j-2}(z_i,z_j|\varepsilon,\infty)\,,\\
    K[h_i,h_j]&\equiv K^{2h_i-2,2h_j-2}(z_i,z_j|\varepsilon,\infty)=\Big(\frac{1}{4\pi\varepsilon}\Big)^3e^{-\frac{|z_{12}|^2}{4\varepsilon}}\Omega_{ij}^{(0,3)}\,,\label{eq:K}\\
    \A[h_i]&\equiv \A_{2h_i-2}(z_i)\,,
\end{align}
\end{subequations}
where $\Omega^{(0,3)}_{ij}=d^3\bar{z}_{ij}\equiv \epsilon_{abc}d\bar{z}_{ij}^ad\bar{z}_{ij}^bd\bar{z}_{ij}^c$. The wheel diagram is given by 
\small
\begin{align}\label{eq:Box}
    &(Fig. \,\ref{fig:anomaly})=\frac{1}{2^4} \int_{(\C^3)^4}\prod_{j=1}^4(d^3 z_j)^4 \times\sum_{\{h'\}}\Tr\Bigg(\nn\\
    &\times\Big( K[h_4',-h_1']\star\A[h_1]\Big)\Big(P[h_1',-h_2']\star \A[h_2]\Big)\Big(P[h_2',-h_3']\star \A[h_3]\Big)\Big( P[h_3',-h_4'] \star\A[h_4]\Big)\Bigg) \,.
\end{align}
\normalsize
Note that there is an explicit kernel $K[h_4',-h_1']$ due to the linearized BRST variation $\delta\A^{p}=\bar{\p}\A^{p-1}$ of one of the BV field.\footnote{Recall that $\bar{\p}P(\epsilon,\infty)=-K(\epsilon,\infty)$, cf. \eqref{eq:bar-p-P} and \eqref{eq:K}.} Furthermore, the symmetry factor for the wheel diagram is $1/2^4$. 

Now, expanding the $\star$-product, we can organize the above as
\begin{align}
    (Fig. \,\ref{fig:anomaly})&=\frac{1}{2^4}  \int_{(\C^3)^4}\prod_{i=1}^4(d^3 z_i)^4\sum_{\{h'\}}\Tr\Bigg[
    \frac{1}{\Gamma[h_4'-h_1'+h_1]}\Pi^{h_4'-h_1'+h_1-1}\Big(K[h_4',-h_1'],\A[h_4]\Big)\nn\\
    &\qquad \qquad \times\frac{1}{\Gamma[h_1'-h_2'+h_2]}\Pi^{h_1'-h_2'+h_2-1}\Big(P[h_1',-h_2'],\A[h_1]\Big)\nn\\
    &\qquad \qquad \times\frac{1}{\Gamma[h_2'-h_3'+h_3]}\Pi^{h_2'-h_3'+h_3-1}\Big(P[h_2',-h_3'],\A[h_2]\Big)\nn\\
    &\qquad \qquad \times\frac{1}{\Gamma[h_3'-h_4'+h_4]}\Pi^{h_3'-h_4'+h_4-1}\Big(P[h_3',-h_4'],\A[h_3]\Big)\Bigg]\,.
\end{align}
From simple counting, we see that \eqref{eq:Box} has $d^9 \bar z$ factors coming from the propagators and the kernel $K$. Using 
\begin{align}
    d\bar{z}_{41}=-(d\bar{z}_{12}+d\bar{z}_{23}+d\bar{z}_{34})\,,
\end{align}
we can write
\begin{align}
    \Big(\prod_{i=1}^3\Omega^{(0,2)}_{i\,i+1}\Big)\Omega^{(0,3)}_{41}
    =-8\epsilon^{a_1a_2a_3}\bar{z}_{12;a_1}\bar z_{23;a_2}\bar z_{34;a_3}d^3\bar z_{12}d^3\bar z_{23} d^3\bar z_{34}\,.
\end{align}
At this stage, it is convenient to make the following change of variables 
\begin{align}
    w_i:=z_{i\,i+1}=z_i-z_{i+1}\,,\quad \bar{w}_i:=\bar{z}_{i\,i+1}=\bar{z}_i-\bar{z}_{i+1}\,
\end{align}
where we note that 
\begin{align}
    \sum_{i=1}^4w_i=0\,,\quad \sum_{j=1}^4\bar{w}_j=0\,.
\end{align}
Then, we can rewrite
\begin{align}
    \Big(\prod_{i=1}^3\Omega^{(0,2)}_{i\,i+1}\Big)\Omega^{(0,3)}_{41}=-8\epsilon^{a_1a_2a_3}\bar{w}_{1\,a_1}\bar{w}_{2\,a_2}\bar{w}_{3\,a_3} \prod_{i=1}^3 d^3\bar{w}_i\,,\qquad a_i=1,2,3\,.
\end{align}
With this, the wheel integral can be written as
\small
\begin{align}
    (Fig. \,\ref{fig:anomaly})&=\frac{1}{2}\sum_{\{h'\}}\cC_{\{h'\}|h}\int_{(\C^3)^4}\prod_{i=1}^3(d^3z_i)^4 \Tr\Big(\prod_{j=1}^{4}\p^{\dot\alpha_j(h'_{j}-h_{j+1}'+h_{j+1}-1)}\A[h_j]\Big)\epsilon^{a_1a_2a_3}\bar{w}_{1\,a_1}\bar{w}_{2\,a_2}\bar{w}_{3\,a_3} \Big(\prod_{t=1}^3 d^3\bar{w}_t\Big)\nn\\
    &\times\int_{\varepsilon}^{\infty} \frac{d\vec{\ell}}{\varepsilon\ell_1\ell_2\ell_3}\,\Big(\prod_{m=1}^3\p_{\dot\alpha_m(h_{m}'-h_{m+1}'+h_{m+1}-1)}k_{\ell_m}(w_m)\Big)\p_{\dot\alpha_1(h_4'-h_1'+h_1-1)}k_{\epsilon}(w_4)
\end{align}
\normalsize
where 
\begin{align}
    d\vec{\ell}=d\ell_1 d\ell_2 d\ell_3\,,\qquad k_{\ell}(w)=\Big(\frac{1}{4\pi\ell}\Big)^3\exp\Big(-\frac{|w|^2}{4\ell}\Big)\,.
\end{align}
In the above,
\begin{align}
    \cC_{\{h'\}|h}=\frac{1}{\Gamma(h_1'-h_2'+h_2)\Gamma(h_2'-h_3'+h_3)\Gamma(h_3'-h_4'+h_4)\Gamma(h_4'-h_1'+h_1)}\,,
\end{align}
is a shorthand notation for the product of coupling constants, which depend on the helicities of fields entering the vertices. Since $\p_{\dot\alpha} k_{\ell}(w)=-\frac{\bar{w}_{\dot\alpha}}{4\ell}k_{\ell}(w)$, we have
\begin{align}
    (Fig. \,\ref{fig:anomaly})&=\frac{1}{2}\sum_{\{h'\}}\cC_{\{h'\}|h}\int_{(\C^3)^4}\Big(\prod_{i=1}^4d^3w_i\Big) 
    \Big(\prod_t d^3\bar{w}_t\Big)\epsilon^{a_1a_2a_3}\bar{w}_{1\,a_1}\bar{w}_{2\,a_2}\bar{w}_{3\,a_3}\nn\\
    &\times\Tr\Big(\prod_{j=1}^{4}(\bar{w}_{j})_{\dot\alpha_j(h'_{j}-h_{j+1}'+h_{j+1}-1)}(\p^{\dot\alpha_j(h'_{j}-h_{j+1}'+h_{j+1}-1)}\A[h_{j}]\Big) \nn\\
    &\times \int_{\varepsilon}^{\infty}\frac{d\vec{\ell}}{\varepsilon\ell_1\ell_2\ell_3}\frac{(-)^{h_1+h_2+h_3+h_4-4}k_{\ell_1}(w_1)k_{\ell_2}(w_2)k_{\ell_3}(w_3)k_{\varepsilon}(w_4)}{(4\varepsilon)^{h_4'-h_1'+h_1-1}(4\ell_ 1)^{h_1'-h_2'+h_2-1}(4\ell_2)^{h_2'-h_3'+h_3-1}(4\ell_3)^{h_3'-h_4'+h_1-1}}\,.
\end{align}
It is now useful to collect all the heat kernels as
\begin{align}
    \frac{k_{\ell_1}(w_1)k_{\ell_2}(w_2)k_{\ell_3}(w_3)k_{\varepsilon}(w_4)}{\varepsilon \ell_1\ell_2\ell_3}=\frac{1}{(4\pi)^{12}}\frac{1}{(\varepsilon\ell_1\ell_2\ell_3)^4}\exp\Big(-\frac{1}{4}\overline{W}_aH^a{}_b W^b\Big)\,,
\end{align}
where \cite{williams2020renormalization}
\begin{align}
    \overline{W}_a=(\bar{w}_{1}, \bar{w}_{2}, \bar{z}_{3})\,,\qquad W^b=\begin{pmatrix}
      w_{1}  \\
       w_{2} \\
       w_{3}
    \end{pmatrix}\,,\qquad H^a{}_b=\begin{pmatrix}
        \frac{1}{\ell_1}+\frac{1}{\epsilon} & \frac{1}{\epsilon} & \frac{1}{\epsilon} \\
        \frac{1}{\epsilon} & \frac{1}{\ell_2}+\frac{1}{\epsilon} & \frac{1}{\epsilon}\\
        \frac{1}{\epsilon} & \frac{1}{\epsilon} & \frac{1}{\ell_1}+ \frac{1}{\epsilon}
    \end{pmatrix}\,.
\end{align}
Then, we can cast the wheel diagram into the form
\begin{align}
    (Fig. \,\ref{fig:anomaly})&=\frac{1}{2(4\pi)^{12}}\sum_{\{h'\}}\cC_{\{h'\}|h}\int_{(\C^3)^4}\Big(\prod_{i=1}^4d^3w_i \Big)\Big(\prod_{t=1}^3d^3\bar{w}_t\Big)\epsilon^{a_1a_2a_3}\bar{w}_{1\,a_1}\bar{w}_{2\,a_2}\bar{w}_{3\,a_3}\nn\\
    &\times\Tr\Big(\prod_{j=1}^{4}(\bar{w}_{j})_{\dot\alpha_j(h'_{j}-h_{j+1}'+h_{j+1}-1)}(\p^{\dot\alpha_j(h'_{j}-h_{j+1}'+h_{j+1}-1)}\A[h_j]\Big) \nn\\
    &\times \int_{\varepsilon}^{\infty}\frac{d\vec{\ell}}{(\varepsilon\ell_1\ell_2\ell_3)^4}\frac{(-)^{h_1+h_2+h_3+h_4-4}\exp\big(-\frac{1}{4}\overline{W}_aH^a{}_bW^b\big)}{(4\varepsilon)^{h_4'-h_1'+h_1-1}(4\ell_1)^{h_1'-h_2'+h_2-1}(4\ell_2)^{h_2'-h_3'+h_3-1}(4\ell_3)^{h_3'-h_4'+h_1-1}}\,.
\end{align}
To continue, we follow \cite{williams2020renormalization} and introduce the differentials
\begin{align}
    \eth_{i\dot\alpha}:=\frac{\p}{\p w^{i\dot\alpha}}+\eth_{4\dot\alpha}\ \ (i=1,2,3)\,,\qquad \eth_{4\dot\alpha}:=-\frac{1}{\epsilon+\ell_1+\ell_2+\ell_3}\sum_{j=1}^3\ell_j\frac{\p}{\p w^{j\dot\alpha}}\,,
\end{align}
where in terms of $z_i$ coordinates \cite{Bittleston:2022nfr}:
\begin{subequations}
    \begin{align}
        \eth_{1 \dot\alpha_1}&=\frac{1}{\varepsilon+\ell_1+\ell_2+\ell_3}\Big[\varepsilon\p_{1\dot\alpha}-(\ell_2+\ell_3)\p_{2\dot\alpha_1}-\ell_3\p_{3\dot\alpha_1}\Big]\,,\\
        \eth_{2 \dot\alpha_2}&=\frac{1}{\varepsilon+\ell_1+\ell_2+\ell_3}\Big[\varepsilon\p_{1\dot\alpha_2}+(\varepsilon+\ell_1)\p_{2\dot\alpha_2}-\ell_3\p_{3\dot\alpha_2}\Big]\,,\\
        \eth_{3 \dot\alpha_3}&=\frac{1}{\varepsilon+\ell_1+\ell_2+\ell_3}\Big[\varepsilon\p_{1\dot\alpha_3}+(\varepsilon+\ell_1)\p_{2\dot\alpha_3}+(\varepsilon+\ell_1+\ell_2)\p_{3\dot\alpha_3}\Big]\,,\\
        \eth_{4 \dot\alpha_4}&=\frac{1}{\varepsilon+\ell_1+\ell_2+\ell_3}\Big[-(\ell_1+\ell_2+\ell_3)\p_{1\dot\alpha_4}-(\ell_2+\ell_3)\p_{2\dot\alpha_4}-\ell_3\p_{3\dot\alpha_4}\Big]\,.
    \end{align}
\end{subequations}
The $\eth$'s derivatives have the following properties 
\begin{subequations}
\begin{align}
    \eth_{4\dot\alpha}\exp\Big(-\frac{1}{4}\overline{W}_aH^a{}_bW^b\Big)&=-\frac{\bar w_{4\dot\alpha}}{4\varepsilon}\exp\Big(-\frac{1}{4}\overline{W}_aH^a{}_bW^b\Big)\,,\\ \eth_{j\dot\alpha}\exp\Big(-\frac{1}{4}\overline{W}_aH^a{}_bW^b\Big)&=-\frac{\bar w_{j\dot\alpha}}{4\ell_j}\exp\Big(-\frac{1}{4}\overline{W}_aH^a{}_bW^b\Big)\,,\qquad j=1,2,3\,.
\end{align}
\end{subequations}
This allows us to replace $\bar w_{1a_1}\bar w_{2a_2}\bar w_{3a_3}$ by $4^3 \ell_1\ell_2\ell_3\eth_{1a_1}\eth_{2a_2}\eth_{3a_3}$. By assuming that $\eth_{ja}$ wrt. to $w_k$ are taken at $z_4$, we can write
\begin{align}
    \epsilon^{a_1a_2a_3}\eth_{1a_1}\eth_{2a_2}\eth_{3a_3}=\frac{\varepsilon}{\varepsilon+\ell_1+\ell_2+\ell_3}\epsilon^{a_1a_2a_3}\p_{1a_1}\p_{2a_2}\p_{3a_3}\,,
\end{align}
where we recall that $\p_i$ are derivatives wrt. $z_i$.

At this stage, we can trade $\bar{w}$'s
for $\eth$'s and do integration by part to obtain
\small
\begin{align}
    (Fig. \,\ref{fig:anomaly})&=-\frac{4^3}{2(4\pi)^{12}}(-)^{h_1+h_2+h_3+h_4-1}\sum_{\{h'\}}\cC_{\{h'\}|h}\int_{(\C^3)^4}\Big(\prod_{i=1}^4d^3w_i \Big)\Big(\prod_{t=1}^3d^3\bar{w}_t\Big)\epsilon^{a_1a_2a_3}\p_{1\,a_1}\p_{2\,a_2}\p_{3\,a_3}\nn\\
    &\times\Tr\Big(\prod_{j=1}^{4}\big(\eth_{\dot\alpha_j}(\p^{\dot\alpha_j}\big)^{h'_{j}-h_{j+1}'+h_{j+1}-1}\A[h_j]\Big)  \int_{\epsilon}^{\infty}\frac{d\vec{\ell}\,\varepsilon}{(\varepsilon\ell_1\ell_2\ell_3)^3}\frac{\exp\big(-\frac{1}{4}\overline{W}_aH^a{}_bW^b\big)}{\varepsilon+\ell_1+\ell_2+\ell_3}\,.
\end{align}
\normalsize
Summing over helicities yields
\begin{align}\label{eq:pre-simplify}
    (Fig. \,\ref{fig:anomaly})&=-\frac{4^3}{2(4\pi)^{12}}\sum_{h'}\frac{(-)^{\tH_4-1}}{\Gamma(\tH_4-3)}\int_{(\C^3)^4}\Big(\prod_{i=1}^3d^3w_i \Big)\Big(\prod_{t=1}^3d^3\bar{w}_t\Big)\nn\\
    &\times\Big(\eth_{1\,\dot\alpha_2}\p^{\dot\alpha_1}+\eth_{2\,\dot\alpha_2}\p^{\dot\alpha_2}+\eth_{3\,\dot\alpha_3}\p^{\dot\alpha_3}+\eth_{4\,\dot\alpha_4}\p^{\dot\alpha_4}\Big)^{\tH_4-4}\Tr\Big(\prod_{j=1}^3(\p\A[h_j])\A[h_4]\Big) \nn\\
    &\times\int_{\varepsilon}^{\infty}\frac{d\vec{\ell}\,\varepsilon}{(\varepsilon\ell_1\ell_2\ell_3)^3}\frac{\exp\big(-\frac{1}{4}\overline{W}_aH^a{}_bW^b\big)}{\varepsilon+\ell_1+\ell_2+\ell_3}\,,
\end{align}
where
\begin{align}
    \tH_4=h_1+h_2+h_3+h_4\,
\end{align}
denote the sum over external helicities. Furthermore, we have simplified our expression by using
\begin{align}
    d^3w\epsilon^{a_1a_2a_3}\p_{1\,a_1}\p_{2\,a_2}\p_{3\,a_3}\A[h_1]\A[h_2]\A[h_3]=\prod_{j=1}^3(\p\A[h_j])\,,
\end{align}
where $\p:=dz^{a}/\p z^a$ is the holomorphic differential operator on $\C^3$.

Notice that there is a sum over helicity $\sum_{h'}$, which will be normalized to 0 as discussed previously in Subsection \ref{sec:index}. Therefore, to have a well-define amplitude, we only need to show the integral over position space is finite. Observe that
\begin{align}
    \Big(\eth_{1\,\dot\alpha_2}\p^{\dot\alpha_1}+\eth_{2\,\dot\alpha_2}\p^{\dot\alpha_2}+\eth_{3\,\dot\alpha_3}\p^{\dot\alpha_3}+\eth_{4\,\dot\alpha_4}\p^{\dot\alpha_4}\Big)=-\Big([\p_1\,\p_2]+[\p_1\,\p_3]+[\p_2\,\p_3]\Big)
\end{align}
where $[\p_i\,\p_j]=\p_i^{\dot\alpha}\p_{j\dot\alpha}$, cf. \eqref{eq:star-product}. Then, the whole expression \eqref{eq:pre-simplify} drastically reduces to
\begin{align}
    (Fig. \,\ref{fig:anomaly})&=\frac{4^3}{2(4\pi)^{12}}\sum_{h'}\frac{1}{\Gamma(\tH_4-3)}\int_{(\C^3)^4}\Big(\prod_{i=1}^3d^3w_i \Big)\Big(\prod_{t=1}^3d^3\bar{w}_t\Big)\nn\\
    &\times\Big([\p_1\,\p_2]+[\p_1\,\p_3]+[\p_2\,\p_3]\Big)^{\tH_4-4}\Tr\Big(\prod_{j=1}^3(\p\A[h_j])\A[h_4]\Big) \nn\\
    &\times\int_{\varepsilon}^{\infty}\frac{d\vec{\ell}\,\varepsilon}{(\varepsilon\ell_1\ell_2\ell_3)^3}\frac{\exp\big(-\frac{1}{4}\overline{W}_aH^a{}_bW^b\big)}{\varepsilon+\ell_1+\ell_2+\ell_3}\,,
\end{align}
At this stage, we can do the Gaussian integral over $w,\bar w$, which gives
\begin{align}
   \int_{(\C^3)^3}\Big(\prod_{i=1}^3d^3w_i \Big)\Big(\prod_{t=1}^3d^3\bar{w}_t\Big) \exp\big(-\frac{1}{4}\overline{W}_aH^a{}_bW^b\big)=(4\pi)^9\Big(\frac{\varepsilon\ell_1\ell_2\ell_3}{\varepsilon+\ell_1+\ell_2+\ell_3}\Big)^3\,.
\end{align}
Then, the wheel amplitude simplifies to
\begin{align}\label{eq:polynomial-form-anomaly}
    (Fig. \,\ref{fig:anomaly})=&\frac{4^3}{2(2\pi)^3}\sum_{h'}\frac{1}{\Gamma(\tH_4-3)} \int_{\varepsilon}^{\infty}\frac{d^3\vec{\ell}\,\varepsilon}{(\varepsilon+\ell_1+\ell_2+\ell_3)^4}\times\nn\\
   \times &\int_{\C^3}\big([\p_1\,\p_2]+[\p_1\,\p_3]+[\p_2\,\p_3]\big)^{\tH_4-4}\Tr\Big(\p\A[h_1]\p\A[h_2]\p\A[h_3]\A[h_4]\Big)\,.
\end{align}
Here, the subscripts in $\p_i$'s indicate, which fields they act on. Note that this is simply a convenient way of writing things, since all derivatives must be understood to be evaluated as the same point.

It is now a simple computation to show that the above integral is convergent. Indeed, we can make a change of variables $\ell_i\mapsto\ell_i\varepsilon$ and integrate over the domain $[1,\infty)^3$. The result of $\ell_i$ integrations is $1/24$. Thus, the integral part of the wheel diagram is finite. Reorganizing \eqref{eq:polynomial-form-anomaly}, our result for the anomalies associated to the wheel diagram Fig. \ref{fig:anomaly} is
\begin{align}\label{eq:pre-exponential}
    (Fig. \,\ref{fig:anomaly})= \frac{1}{3!(2\pi)^3}\sum_{h'}\int_{\C^3}\frac{\big([\p_2\,\p_3]+[\p_2\,\p_4]+[\p_3\,\p_4]\big)^{\tH_4-4}}{(\tH_4-4)!}\Tr(\A\p\A_2\p\A_3\p\A_4)\Bigg|_{\substack{\A_{2,3,4}=\A\\
    \p_{2,3,4}\mapsto\p}}\,.
\end{align}
\normalsize
Note that we are left with the sum over all dof., which is regularized to zero, as alluded to in the above, since we are considering $\Spec=\Z$ case.

\paragraph{Remarks.} In our computations, we have considered a spectrum comprising all integer spins, resulting in a rather compact and simple expression \eqref{eq:pre-exponential}. Had we considered theories with $\Spec=2\Z+1$ or $\Spec=2\Z$, the expression \eqref{eq:pre-exponential} would be more intricate due to additional derivative structures. Nonetheless, the proof of convergence for the one-loop integral would remain the same. 

This leads to the conclusion that holomorphic twistorial theories, whether higher-spin or not, are anomalous at one loop if they do not have the ``right'' spectrum. Although the twistorial anomalies in this context are gauge anomalies, it is, however, interesting to consider them as it would suggest that the dual physical spacetime higher-spin theories can have non-trivial scattering amplitudes. 
This correspondence can be summarized as 
\begin{center}
    anomalous twistor theories $\longleftrightarrow$ interesting $4d$ spacetime theories.
\end{center}
Note that twistorial gauge anomalies can be resolved by extending the field content of the anomalous twistorial theories, as shown below. Furthermore, instead of working with holomorphic bosonic theories, we can also generalize them to supersymmetric ones by including suitable fermionic terms. Then, as observed in e.g. \cite{Costello:2019jsy}, the coefficient in front of the loop amplitudes are typically proportional to $N_{boson}-N_{fermion}$. This implies super symmetric theories should have better quantum properties than the bosonic ones, at least up to some number of loops. 
\subsection{Anomaly cancellation mechanisms for holomorphic theories}
This subsection studies quantum properties of all possible higher-spin theories that can be derived from the BV actions \eqref{eq:BV-action}, cf. Table \ref{tab:a}, and Table \ref{tab:b}. In particular, we generalize the methods in \cite{Costello:2015xsa,Costello:2021bah,Bittleston:2022nfr} to render anomalous higher-spin theories quantum consistent at one loop. 

\subsubsection{Descendant higher-spin theories} 

As it is well-known, in the absence of interactions, the kinetic action part of \eqref{eq:kinetic-action} can serve well for any spin, see e.g. \cite{Tran:2021ukl}. However, as soon as interactions involving fields with spin greater than two are introduced, it is well-known that one must include a tower of infinitely many massless higher-spin fields to ensure the classical consistency of deformed higher-spin theories. Note that while this technical result applies to spins, it does not mean that number of the derivatives in a vertex should go to infinite as well. In particular, for the holomorphic twistorial theories in consideration, there are possible truncations of \eqref{eq:BV-action} to its closed subsectors with lower-derivative interactions as shown in Section \ref{sec:3}. 

In studying quantum consistency of various theories resulting from the actions \eqref{eq:BV-action} and \eqref{eq:BV-action-BF}, we notice that not all of them are anomaly-free. As usual, we classify these theories into two sectors: (i) colored one; and (ii) colorless one.


\paragraph{Colored theories.} In the colored sector, we observe that hCS${}^{\star}_{\mg}$ (with all spins or odd spins) and HS-BF${}_{\mg}$ (with odd spins only) are self-quantum consistent, while others are anomalous at one loop, cf. Table \ref{tab:a}. In summary, 

\begin{table}[ht!]
    \centering
    \begin{tabular}{|c|c|c|c|}\hline
       Theory  & spectrum & interaction type & quantum consistency  \\ \hline\hline
       $\hCS_{\mg}$ & all & $\star$ & $\checkmark$ \\\hline
        $\hCS_{\mg}$ & all/odd spins & $ [\,,]_{\mg} $ & $\checkmark$ \\\hline
       HS-BF${}_{\mg}$ & $s\geq 1$ & $ [\,,]_{\mg}$ & $\boldsymbol{\times}$ \\\hline
       HS-BF${}_{\mg}$ & odd spins & $ [\,,]_{\mg}$ & $\checkmark$ \\\hline
    BF${}_{\mg}$ & $s=1$ & $[\,,]_{\mg}$ & $\boldsymbol{\times}$ \\\hline
    \end{tabular}
    \caption{Table of descendant \underline{\emph{colored}} higher-spin theories derived from the parent actions \eqref{eq:parent-action-1} and \eqref{eq:BF-parents}. The last column indicates  their quantum properties at one loop.} 
    \label{tab:1}
\end{table}

\paragraph{Colorless theories.} For the colorless sector cf. Table \ref{tab:b}, we have
\begin{table}[ht!]
    \centering
    \begin{tabular}{|c|c|c|c|}\hline
       Theory  & spectrum & interaction type & quantum consistency  \\ \hline\hline
       $\hCS$ & all & $\star$ & $\checkmark$ \\\hline
       $\hCS$ & all/(even+0) & $ \{\,,\} $ & $\checkmark$ \\\hline
       HS-BF${}_{GR}$ & $s\geq 2\,;\ s\in 2\N$ & $ \{\,,\}$ & $\boldsymbol{\times}$ \\\hline
       BF${}_{GR}$ & $s=2$ & $\{\,,\}$ & $\boldsymbol{\times}$ \\\hline
    \end{tabular}
    \caption{Table of descendant \underline{\emph{colorless}} higher-spin theories derived from the parent actions \eqref{eq:parent-action-1} and \eqref{eq:BF-parents}. The last column indicates  their quantum properties at one loop.}
    \label{tab:2}
\end{table}

\subsubsection{Anomaly canceling mechanisms} 
From the tables above it is clear that except higher-spin self-dual Yang-Mills with only odd spins in the spectrum
, all other self-dual theories are anomalous at one loop. Thus, it is natural to ask how to render these theories quantum consistent, as the anomalies we are facing are gauge anomalies, which can be fatal if they lead to inconsistency in the corresponding spacetime theories. (Note that theories with $\Spec=\{|h|\geq 2\}$ or $\Spec=\{|h|\in2\N\}$ appear to be more complicated than those whose spectrum has no spin gap at one loop.) 

For low-spin holomorphic theories with $s\leq 2$, this question has been addressed in \cite{Costello:2021bah,Bittleston:2022nfr}. Here, we extend their analysis to higher-spin cases.\footnote{Note that some spacetime counterpart of our discussion can be found in \cite{Monteiro:2022xwq}.} It is useful to point out the essence of the Green-Schwarz-like anomaly cancellation mechanism in \cite{Costello:2015xsa,Costello:2021bah,Bittleston:2022nfr} as follows. Since the wheel diagram, Fig. \ref{fig:anomaly}, associated with the gauge anomaly on twistor space is of the form $\int \cc(\partial \sA)^3$, it can be cancelled by considering a suitable 4-point tree-level amplitude with some fine-tuned couplings, whose linearized BRST (gauge) transformation reproduces the exact anomaly with the opposite sign. We shall demonstrate this \emph{on-shell} statement below.

\paragraph{HS-BF${}_{\mg}$.} Let us start with the anomaly canceling mechanism for HS-BF${}_{\mg}$ theory with $\Spec=\{|h|\in\N\}$, which is one-loop exact, similar to other BF-type theories. Note that there is no spin-0 running in the loop, and the sum over the spectrum gives $2\sum_{s\geq 1}1=-1$ (using Riemann zeta regularization) for the anomaly.\footnote{Recall that  the 2 stands for the degrees of freedom of massless spinning fields in $4d$.} 

Since the cubic vertices are of type $\sA_{-2s_1-2}[\sA_{2s_2-2},\sA_{2s_3-3}]$, where $s_2+s_3-s_1=1$ and $s_{1,2,3}\geq 1$, the external fields of the one-loop amplitudes in this case must be $+1$. Translating the above information to the wheel diagram, cf. Fig \ref{fig:anomaly}, we see that all external physical gauge fields must be $\sA_0$ for the case of HS-BF${}_{\mg}^{[,]}$. This gives\footnote{This can also be confirmed by redoing the computation in Subsection \ref{sec:loop-computation} with all higher-derivative terms removed.}
\begin{align}\label{eq:loop-HS-BF}
    \Ccurl_{\text{HS-BF}_{\mg}}^{[,]}\Big(\parbox{30pt}{\includegraphics[scale=0.06]{anomaly.png}}\Big)=-\frac{i^3}{3!(2\pi)^3}\int_{\PT}\Tr\Big(\cc\wedge(\p \sA_{0})\wedge (\p\sA_0)\wedge(\p\sA_{0})\Big)=-\Ccurl_{\text{BF}_{\mg}}\,,
\end{align}
where we have set $\tH_4=4$, and $\Ccurl_{\text{BF}_{\mg}}$ was computed in e.g. \cite{Costello:2021bah,Bittleston:2022nfr}. Therefore, the gauge anomaly of HS-BF${}_{\mg}$ is negative of that of the usual BF${}_{\mg}$ theory on twistor space. 

There are two possible way to cure the above anomaly. First, we can consider the following quantum corrected term
\begin{align}
    \int_{\PT}\Tr\Big(\varphi_{-2}\bar{\p}\varphi_{-2}+\varphi_{-2}[\sA_0,\varphi_{-2}]\Big)\,,\qquad \varphi_{-2}\in \Omega^{0,1}(\PT,\cO(-2)\otimes \mg)\,,
\end{align}
where $\varphi$ is a scalar field taking valued in the adjoint. This allows the spin-0 field to run in the loop, thus trivializing the sum over the spectrum in \eqref{eq:polynomial-form-anomaly}. This theory may be identified with the 1-derivative chiral theory found in \cite{Ponomarev:2017nrr}.


The second method for anomaly cancellation (adapted to higher-spin setting) follows from the recent Green-Schwarz type anomaly cancellation in twistor space \cite{Costello:2021bah}. It goes as follows. 

First, as observed by Okubo, cf. \cite{Okubo:1978qe}, there is an intriguing relationship 
\begin{align}\label{eq:Okubo}
    \Tr(T^{(a_1}T^{a_2}T^{a_3} T^{a_4)})=C_{\mg}\tr(T^{(a_1}T^{a_2})\tr(T^{a_3}T^{a_4)})\,, 
\end{align}
between the trace in the adjoint representation (denoted as $\Tr$) and the double trace in the fundamental representations (denoted as $\tr$).\footnote{The matrices $T^a$ are assumed to be in the appropriate representations for evaluating such traces.} Here, $T^a$ are the generators of one of the Lie algebra of $SU(2),SU(3),SO(8)$, or of the exceptional group $E_{5,6,7}$. The term $C_{\mg}$ reads
\begin{align}
    C_{\mg}=\frac{10 h^{\vee}}{2+\dim(\mg)}\,,
\end{align}
where $h^{\vee}$ is the Coxeter number of the aforemention Lie algebras. Using the Okubo's relation, cf. \eqref{eq:Okubo}, we write
\begin{align}\label{eq:Tr-tr}
    \Tr\Big(\A[h_1]\p\A[h_2]\p\A[h_3]\p\A[h_4]\Big)=C_{\mg}\tr\Big(\A[h_1]\p\A[h_2]\Big)\tr\Big(\p\A[h_3]\p\A[h_4]\Big)\,.
\end{align}
Our task is to construct additional interactions with some new fields in the fundamental representations of $\mg$ such that the linearized BRST transformations of tree-level diagrams with the new fields in the exchange result in the rhs. of \eqref{eq:Tr-tr}. Then, upon fine-tune the coupling constant $C_{\mg}$ we can cancel the (unwanted) anomaly \eqref{eq:loop-HS-BF}.

Consider the following axionic quantum corrected BV action (inspired by \cite{Costello:2021bah})
\begin{align}\label{eq:axion-BVaction}
    S^{cor,BV}_{\text{HS-BF}_{\mg}}=\int_{\PT}\p^{-1}\boldsymbol{\vartheta}\bar{\p}\boldsymbol{\vartheta}+c_{\mg}\int_{\PT}\boldsymbol{\vartheta}\tr(\A_0\p\A_0)\,,\qquad \boldsymbol{\vartheta}\in \Omega^{2,\bullet}(\PT,\cO(0))[1]\,,
\end{align}
whose tree-level diagram associated with the vertex $\int \boldsymbol{\vartheta}\tr(\A_0\p\A_0)$ that involves the axion field running in the exchange can cancel the one-loop amplitude \eqref{eq:loop-HS-BF}, 
\begin{align}\label{eq:total-axion-loop}
    \parbox{95pt}{\includegraphics[scale=0.2]{anomaly.png}}+\parbox{95pt}{\includegraphics[scale=0.23]{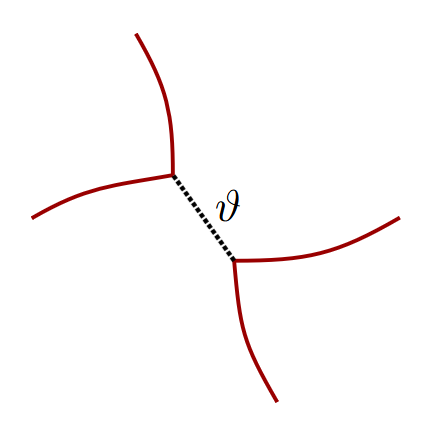}}=0\,.
\end{align}
In eq. \eqref{eq:axion-BVaction}, $\p^{-1}:\Omega^{p,\bullet}(\PT)\rightarrow \Omega^{p-1,\bullet}(\PT)$ is the formal inversion of the holomorphic differential $\p:=dz^a\p_a$ for $a=1,2,3$, and $\boldsymbol{\vartheta}$ is a BV axion field whose components are
\begin{align}
    \boldsymbol{\vartheta}=\chi^{2,0}+\vartheta^{2,1}+\vartheta^{2,2}_{\vee}+\chi^{2,3}_{\vee}\,.
\end{align}
As usual, $\chi,\chi_{\vee}$ are the ghost and antifield of ghost, while $\vartheta,\vartheta_{\vee}$ are field and antifield in BV language. The form degrees of the corresponding fields are evident from the way we index them. 

The important requirement for the physical axion $\vartheta^{2,1}$ is that it should obey $\p\vartheta=0$. Note that this axion field can be viewed as the Beltrami differential, which is a vector-valued $(0,1)$-form, i.e. element of $\Omega^{0,1}(\PT,T^{1,0}\PT)$. The field $\vartheta^{2,1}$ transforms as
\begin{align}
    \delta \vartheta^{2,1}=\bar{\p}\varpi^{2,0}\,,\qquad \varpi^{2,0}\in \Omega^{2,0}(\PT,\mg)\,.
\end{align}
In trivializing the total amplitude \eqref{eq:total-axion-loop}, we can set
\begin{align}
    c_{\mg}=\sqrt{\frac{-iC_{\mg}}{3!(2\pi)^3}}\,.
\end{align}
A sketch of the anomaly cancellation via the Green-Schwarz-like mechanism can be found in Appendix \ref{app:B}.

\paragraph{HS-BF${}_{GR}$.} The anomaly-canceling mechanism for this sector is almost the same as in the HS-BF${}_{\mg}$ case, with a few adjustments. First, the sum over $\Spec=2\Z^*$ is again evaluated to $2\sum_{s\in 2\N}1=-1$, while for $\Spec=\{|h|\geq 2\}$, we get $2\sum_{s=2}^{\infty}1=-3$. Second, the lowest-derivative interaction term is the Poisson bracket $\{\sA,\sA\}$. Then, we see that the external fields of the one-loop amplitudes in this case must be $+2$ based on the spin constraint $$s_2+s_3-s_1=2\,,$$
    where $s_{1,2,3}\geq 2$. Here, we consider only the case with $\Spec=\{|h|\geq 2\}$ while leaving the case $\Spec=2\Z^*$ as an exercise. 
    
    The wheel diagram for HS-BF${}_{GR}$ with $\Spec=\{|h|\geq 2\}$ is evaluated to
    \small
\begin{align}
    \Ccurl_{\text{HS-BF}_{GR}}^{\Spec=\{|h|\geq 2\}}=\frac{i}{3!(2\pi)^3}\int_{\C^3}\frac{\big([\p_1\,\p_2]+[\p_3\,\p_4]\big)^{4}}{4!}\Big(\cc_2(z_1)\p \sA_{2}(z_2)\p \sA_{2}(z_3)\p \sA_{2}(z_4)\Big)\Big|_{z_i=z}\,.
\end{align}
\normalsize
Then, our result for the anomaly of BF${}_{GR}$ is
\begin{align}
    \Ccurl_{\text{BF}_{GR}}= \frac{i^3}{ 3!4!(2\pi)^3}\int_{\C^3}\big([\p_1\,\p_2]+[\p_3\,\p_4]\big)^{4}\Big(\cc_2(z_1)\p \sA_{2}(z_2)\p \sA_{2}(z_3)\p \sA_{2}(z_4)\Big)\Big|_{z_i=z}\,,
\end{align}
where we have used the fact that $2\sum_{s\in 2\N}1=-1$.

Again, there are two ways to cure the anomaly associated with HS-BF${}_{GR}$. Namely, we can either introduce the spin-0 and spin-1 fields with appropriate couplings into the system, or we can introduce an axionic quantum corrected BV action on $S^7$
\begin{align}
   S_{\text{HS-BF}_{GR}}^{cor,BV}= \int_{S^7}  \p^{-1}\varthetabold\bar{\p}\varthetabold+c_{GR}\int_{S^7} \varthetabold \A_2\star \p \A_2\,,\qquad \varthetabold\in H^{2,1}(\PT,\cO(0))[1]
\end{align}
to cancel the loop amplitude using the tree-level $\vartheta$-exchange diagram.\footnote{As shown many times in this work, the integration over the auxiliary $U(1)$ manifold will select the correct interactions on the nose.} See Appendix \ref{app:B} for a short illustration. Note that in the former mechanism, the presence of the spin-0 and spin-1 fields in the spectrum necessarily morph HS-BF${}_{GR}$ into $\hCS_{GR}^{\{\,,\}}$ (see Table \ref{tab:2}) with $\{,\}$-type interactions. As observed in \cite{Bittleston:2022nfr}, this theory is self-quantum consistent, i.e. anomaly-free.



\section{Discussion}\label{sec:discuss}
There is a growing body of evidence suggesting that, it may be necessary to incorporate higher-spin fields (either massless or massive) and higher-derivative interactions in various gravitational theories to ensure their UV completeness at high energy limit.\footnote{These higher-derivative interactions, often referred to as irrelevant deformations, have negligible effect at low energies and can generally be ignored in the infrared (IR) regime.} As a side effect, this also helps to maintain Lorentz invariance and support symmetry enhancements at the high energy regime. Therefore, the study of higher-spin theories, as reviewed in e.g. \cite{Bekaert:2022poo}, could serve as useful toy models for exploring fundamental aspects of quantum gravity, at least within the perturbative regime.

As is common in the Lagrangian approach to quantum field theory, a complicated spacetime effective action of a certain theory sometimes can be expressed more elegantly by formulating the same theory in another auxiliary space, with the worldsheet of superstring theory is a prime example. In fact, it is often that the top-down formulation of the desired theory in the auxiliary space can be more natural and respects locality up to a certain degree -- i.e. interactions between fields do not contain $\Box^{-1}$ or some formal divergence sums of the form $\sum_n\Box^n$. (See also \cite{Metsaev:2005ar,Metsaev:2025qkr} for discussions of locality in the light-cone gauge.) Note that this does not necessarily imply that the spacetime description of the same theory will be local when we integrate out some of the auxiliary fields from the spectrum.



Since a worldsheet description for higher-spin theories is not yet available,\footnote{Another interesting approach is to work with holographic collective dipoles (see, e.g., \cite{deMelloKoch:2014vnt,deMelloKoch:2018ivk}), where the dual gravitational theory is defined via a bi-local field theory on the CFT side.} one of the closest alternative spaces for handling potential spacetime non-locality is twistor space. Due to its deep connection with integrability, most theories constructible from twistor space naturally inherit classical integrability, see e.g. \cite{Haehnel:2016mlb,Adamo:2016ple,Krasnov:2021nsq,Tran:2021ukl,Herfray:2022prf,Adamo:2022lah,Tran:2022tft,Neiman:2024vit}.

The proposed BV actions in this work aims to unify a large class of classical integrable holomorphic theories with vanishing cosmological constant.\footnote{See also \cite{Mason:2025pbz} for complimentary results.} Additionally, we investigated the quantum consistency of various holomorphic higher-spin theories on twistor space whose spacetime duals are expected to be chiral/self-dual higher-spin theories, cf. \cite{Metsaev:1991mt,Metsaev:1991nb,Ponomarev:2016lrm,Ponomarev:2017nrr,Metsaev:2018xip,Metsaev:2019dqt,Metsaev:2019aig,Tsulaia:2022csz,Sharapov:2022wpz,Sharapov:2022awp}. 
Our result implies that the one-loop amplitudes of the $4d$ higher-spin theories, whose twistor duals are anomalous (before introducing the axion fields), should be nontrivial, thereby avoiding the standard conclusion of no-go theorems \cite{Weinberg:1964ew,Coleman:1967ad} in flat space. 

Similarly to \cite{Monteiro:2022xwq}, we also observed that while HS-BF's theories are one-loop exact, they appear to be more complex than $\hCS^{\star}$ theories at one loop. Furthermore, their one-loop amplitudes can be computed simply from tree-level axion-exchange diagrams. On the other hand, $\hCS^{\star}$ is not one-loop exact. This poses significant challenges in studying $\hCS^{\star}$ at higher loops. (See \cite{Skvortsov:2018jea,Skvortsov:2020wtf,Skvortsov:2020gpn} for the one-loop amplitudes of chiral higher-spin gravity in spacetime). 
Thus, one might consider the recent advancements in \cite{Costello:2022upu,Costello:2023vyy,Bittleston:2024efo} for bootstrapping higher-loop integrands or form factors using the operator product expansion of chiral vertex algebras, which are Koszul dual with the higher-spin symmetries underlying holomorphic theories in twistor space.

What one finds using the technology in \cite{Costello:2022upu,Costello:2023vyy,Bittleston:2024efo} is that the correlation functions of the chiral vertex algebras can be identified with the $4d$ spacetime rational amplitudes. (These are the loop amplitudes that do not contain UV and IR divergences due to the integrability of the chiral/self-dual sectors.) 
One of the criteria for the framework of \cite{Costello:2022upu,Costello:2023vyy,Dixon:2024tsb,Bittleston:2024efo} to work is to make sure that the partition function of the bulk/defect D${}_5$-D${}_1$ brane system is gauge invariant in all order in perturbation theory.\footnote{The bulk is the twistor space, and the defect is a complex co-dimensional two twistorial sphere.}  
With the inclusion of higher spins, it may introduce subtleties, making the study of chiral algebra associated with holomorphic twistorial higher-spin theories a gratifying problem. Addressing these challenges would enable one to bypass summing over numerous Feynman diagrams by working with elegant algebraic structures of the chiral algebra. We should emphasize that this bootstrapping method only works for specific theories with strong integrability properties such that their amplitudes are rational.

Lastly, we note that even though most $4d$ higher-spin theories are classically integrable in the sense of Bardeen \cite{Bardeen:1995gk}, as evidenced by the studies of tree-level scattering amplitudes in e.g. \cite{Joung:2015eny,Beccaria:2016syk,Roiban:2017iqg,Tran:2022amg}, there are exceptions such as higher-spin Yang-Mills \cite{Adamo:2022lah} and its higher-dimensional counterpart \cite{Basile:2024raj}, where classical integrability may not hold. This raises the intriguing possibility of examining these theories more closely to understand how they navigate the challenges posed by no-go theorems \cite{Weinberg:1964ew,Coleman:1967ad}. We expect that an explicit light-cone analysis within Metsaev’s framework, cf. \cite{Metsaev:2005ar}, or a study of chiral vertex algebras should, in principle, shed light on this matter.


\section*{Acknowledgement}
The author gratefully acknowledges valuable discussions with Tim Adamo, Thomas Basile, Roland Bittleston, Yannick Herfray, Arthur Lipstein, Lionel Mason, Tristan McLoughlin, Ricardo Monteiro, Yasha Neiman and Zhenya Skvortsov on various aspects of twistor theory over the past several years. The author appreciates Tim Adamo, Mattia Serrani and Zhenya Skvortsov for the correspondence. He also thanks Pantelis Panopoulos and Matthew Roberts for various useful discussion. The author is grateful to UMONS for the hospitality during the workshop ``Twistors and Higher Spins" in 2024, Mons, Belgium.  This work is supported by the Young Scientist Training (YST) program at the Asia Pacific Center of Theoretical Physics (APCTP) through the Science and Technology Promotion Fund and Lottery Fund of the Korean Government, and also the Korean Local Governments
– Gyeongsangbuk-do Province and Pohang City.


\appendix


\section{Quantum BV formalism \`a la Wilson}\label{app:RGEFT} 
This appendix reviews some key concepts in \cite{costello2022renormalization}, which is a regularized version of the usual BV-BRST formalism (see e.g. \cite{Cattaneo:2019jpn} for an elegant introduction). 

As is well-known, renormalization group (RG) flow allows tracking the behavior of an effective field theory across a range of length scale (the ``rough'' inverse of energy scale), say $\ell\in [\ell_1,\ell_2]$ where $0\leq \ell_1\leq \ell_2< \infty$. This idea is combined with the BV formalism to form a RG-compatible BV formalism in \cite{costello2022renormalization}. 
Denote $S_{BV}^{int}[\ell]$ as the interaction part of a BV action defined within the length scale $\ell\in[\ell_1,\ell_2]$. 
This theory is said to be renormalizable if the corresponding $S_{BV}^{int}[\ell]$ action requires having only a finite number of counter terms in the limit $\ell_1\rightarrow 0$ and $\ell_2\rightarrow \infty$. The behavior of $S_{BV}^{int}[\ell]$ is captured by the Wilsonian flow operator $\cW(\ell_2,\ell_1)$ as
\begin{align}
    S_{BV}^{int}[\ell_2]=\cW_{[\ell_2,\ell_1]}S_{BV}^{int}[\ell_1]\,,\qquad \ell_1\leq \ell_2\,,
\end{align}
where 
\begin{align}
    \cW_{[\ell_2,\ell_1]}S_{BV}^{int}[\ell_1]=\hbar\log\Big(\exp\big[\hbar\Delta(\ell_2,\ell_1)\big]\exp\big[\frac{S_{BV}^{int}[\ell_1]}{\hbar}\big]\Big)\,,
\end{align}
for $\hbar$ a \emph{formal parameter}, whose restriction $\hbar=0$ should map the quantum BV action $S_{BV}[\ell]$ to its classical contribution. The \emph{flow}-differential operator, or rather the regularized Laplacian operator, between the length scale $[\ell_1,\ell_2]$ is given by 
\begin{align}
    \Delta(\ell_2,\ell_1)=\int d^dxd^dy P(x,y|\ell_1,\ell_2)\frac{\delta}{\delta \Phi(x)}\frac{\delta }{\delta \Phi (y)}\,,
\end{align}
where $\Phi(x)$ denote the BV fields and $P(x,y|\ell_1,\ell_2)$ are their corresponding regularized propagators. Let $\cM^d$ be a spacetime manifold of $d$ dimensions. Then, the expression of $P(x,y|\ell_1,\ell_2)$ in position space $\cM^d\times \cM^d$ reads
\begin{align}
    P(x,y|\ell_1,\ell_2)=
    \int_{\ell_1}^{\ell_2}\frac{d\ell}{2\ell}\Big(\frac{1}{4\pi\ell}\Big)^{\frac{d}{2}}e^{-\frac{|x-y|^2}{4\ell}}\equiv 
    \int_{\ell_1}^{\ell_2}\frac{d\ell}{2\ell}\, K(x,y|\ell)\,.
\end{align}
In the above, $K(x,y|\ell)$ is the regularized heat kernel associated with the propagator $P(x,y)$. The reader may notice that $\ell$ is in fact the Schwinger parameter, which corresponds to the proper time. 

The upshot is that we can obtain the regularized quantum master equation (QME):
\begin{align}
   \hbar \Delta_{\ell} S_{BV}^{int}[\ell]+\frac{1}{2}(S_{BV}^{int}[\ell],S_{BV}^{int}[\ell])_\ell=0
\end{align}
by requiring the interacting part of the BV action $\frac{1}{\hbar}S^{int}_{BV}[\ell]$, to be in the cohomology of the regularized Laplacian $\Delta_{\ell}$ for a given range of length scale $\ell\in [\ell_1,\ell_2]$ where $(-,-)_{\ell}$ stands for the regularized BV bracket. 

Now, if the above equation can make sense in the limit $\ell_1\rightarrow 0$ (UV cutoff) and $\ell_2\rightarrow \infty$ (IR cutoff), then the family of effective BV theories $S_{BV}^{int}[\ell]$ is said to be renormalizable. Furthermore, two solutions of QME are homotopic iff they are related by a symplectic change of coordinates. Also, two theories which are homotopic should be viewed as being equivalent, as they differ by a field redefinition.

As is well-known, the above BV-BRST formalism works naturally well with theories formulated in first-order formalism. This is due to the fact that they naturally admit $d^{\dagger}$, which is the formal adjoint of the de Rham operator, as a gauge-fixing operator. This means that $\text{Im}(d^{\dagger})$ can be used to define the isotropic Lagrangian subspace in the space of BV fields. Moreover, homotopies between theories correspond to variations of $d^{\dagger}$ along the RG flow.


The extension of the above concepts to the holomorphic setting, cf. \cite{williams2020renormalization}, where the gauge-fixing operator is $\bar{\p}^{\dagger}:\Omega^{0,k}\rightarrow\Omega^{0,k-1}$ are done in the main text. Note that to ease the notation, we have implicitly set $\hbar=1$ throughout in the main text.
\section{On the Green-Schwarz-like mechanism}\label{app:B}
This appendix provides the detail of the anomaly cancellation via the Green-Schwarz like mechanism. Let us begin with the standard story of HS-BF${}_{\mg}$ with $[\,,]_{\mg}$-type interaction. Recall that the quantum corrected action on twistor space is
\begin{align}
    S^{cor}_{\text{HS-BF}_{\mg}}=\int_{\PT}\p^{-1}\vartheta\bar{\p}\vartheta+c_{\mg}\int_{\PT}\vartheta\tr(\sA_0\p\sA_0)\,,\qquad \vartheta\in \Omega^{2,1}(\PT,\cO(0))\,,
\end{align}
where 
\begin{align}
    \delta \vartheta=\bar{\p}\varpi\,,\qquad \delta \sA=\bar{\p}\cc+[\sA,\cc]\,.
\end{align}
It can be checked that the non-linear terms in the gauge transformation of $\sA$ cancel out thanks to the fact that $\p\vartheta=0$. However, the cubic vertex $\int \vartheta \tr(\sA_0\p\sA_0)$ is only gauge invariant on-shell, where we require $\bar{\p}\sA_0\approx 0$. In what follows, we shall show that the linearized BRST transformation of $\delta\sA_0$ will lead to the anomaly term that can cancel the wheel diagram (Fig. \ref{fig:anomaly}) in the main text.

Recall that the $\vartheta$-propagator, which is formally a $(4,3)$-form, obeying
\begin{align}\label{eq:delta-P-vartheta}
    \lim_{\varepsilon\rightarrow 0}\bar{\p}P_{\vartheta}(z,z'|\varepsilon,\infty)=-\p\delta^{3,3}(z-z')\,,\qquad \p:=dz^a\frac{\p}{\p z^a}\,,\quad a=1,2,3\,,
\end{align}
where $z$ and $z'$ are points on $\PT\times_{\cM}\PT\simeq \PT\times \P^1$. That is, the interactions between vertices $\vartheta\tr(\sA\p\sA)$ can be non-local in $\PT\times\P^1$, 
where $\P^1$ is another holomorphic curve fibered over the same spacetime points $\tp\in \cM$. However, in the $\varepsilon\rightarrow 0$ limit, this interaction must reduce to a local one as to cancel the anomaly on twistor space. 
With this consideration, the axionic tree-level amplitude reads
\begin{align}
 I_{\vartheta}^{[,]_{\mg}}=\lim_{\varepsilon\rightarrow 0} \frac{c^2_{\mg}}{4} \int_{\PT\times_{\cM}\PT}\tr(\sA_0\p\sA_0)P_{\vartheta}(z,z'|\varepsilon,\infty)\tr(\sA_0\p\sA_0)\,,
\end{align}
where $1/4$ is the symmetry factor. Under the linearized gauge transformation $\delta \sA=\bar{\p}\cc$, the above amplitude varies as
\begin{align}
  I_{\vartheta}^{[,]_{\mg}}=   \lim_{\varepsilon\rightarrow 0}c^2_{\mg}\int_{\PT\times_{\cM}\PT}\tr(\bar{\p}\cc_0\p\sA_0)P_{\vartheta}(z,z'|\epsilon,\infty)\tr(\sA_0\p\sA_0)\,.
\end{align}
Since $\bar{\p}\sA_0\approx 0$ (on-shell) and $\p P_{\vartheta}=0$, we can do integration by part and obtain
\begin{align}
  I_{\vartheta}^{[\,,]_{\mg}}= c^2_{\mg} \int_{\PT\times_{\cM}\PT} \tr(\cc_0\p\sA_0)\p\delta^{3,3}(z-z')\tr(\sA_0\p\sA_0)\,,
\end{align}
by virtue of \eqref{eq:delta-P-vartheta}. Then, upon doing another integration by part and evaluating the integral on the support of $\delta^{3,3}(z-z')$, 
\begin{align}
   I_{\vartheta}^{[\,,]_{\mg}}=  -c^2_{\mg}\int_{\PT}\tr(\cc_0\p\sA_0)\tr(\p\sA_0\p\sA_0)\,,
\end{align}
which cancels with the anomaly \eqref{eq:loop-HS-BF} using Okubo's relation, cf. \eqref{eq:Okubo}.

Moving on to the case of HS-BF${}_{GR}$, where the interaction has the following heuristically form $\int_{S^7}\vartheta(\sA\star\p\sA)$, we observe a similar pattern for anomaly cancellation. In particular, the axionic tree-level amplitude in this case is
\begin{align}
 I_{\vartheta}^{\{,\}} = \lim_{\varepsilon\rightarrow 0} \frac{c^2_{GR}}{4} \int_{\PT\times_{\cM}\PT}(\sA\star \p\sA)P_{\vartheta}(z,z'|\varepsilon,\infty)(\sA\star\p\sA)\,.
\end{align}
By considering the linearized gauge transformation $\delta\sA=\bar\p \cc$ and doing integration by part appropriately, we obtain
\begin{align}
   I_{\vartheta}^{\{,\}}= -c^2_{GR}\int_{S^7}(\cc\star\p\sA)(\p\sA\star\p\sA)\,,
\end{align}
Upon doing an appropriate pushforward to the base $\PTc$, we can get anomaly cancellation by fine-tuning the coupling constant $c_{GR}$. 





\footnotesize
\bibliography{twistor.bib}
\bibliographystyle{JHEP-2}

\end{document}